\newif{\ifarxiv}
\newif{\ifdraft}
\newif{\ifremarks}

\arxivtrue 
\drafttrue 
\remarkstrue 

\ifdraft
\else
\fi

\documentclass[12pt,a4paper]{article}

\usepackage[utf8]{inputenc}
\usepackage{amsmath}
\usepackage[dvipsnames,table]{xcolor}
\usepackage{cite}
\usepackage{subfigure}

\usepackage{listings}

\definecolor{mathgreen}{RGB}{0,90,39}
\ifremarks
\newcommand{\remarktb}[1]{{\renewcommand{\bfdefault}{b}\color[RGB]{0,150,0}{\textbf{T:~#1}}}}
\newcommand{\comP}[1]{{\bf\textcolor{magenta}{P:~#1}}}
\newcommand{\comA}[1]{{\bf\textcolor{blue}{A:~#1}}}
\newcommand{\comD}[1]{{\bf\textcolor{red}{D:~#1}}}
\newcommand{\comC}[1]{{\bf\textcolor{teal}{C:~#1}}}
\fi
\providecommand{\remarktb}[1]{\ignorespaces}
\providecommand{\comP}[1]{\ignorespaces}
\providecommand{\comA}[1]{\ignorespaces}
\providecommand{\comD}[1]{\ignorespaces}
\providecommand{\comC}[1]{\ignorespaces}

\usepackage[T1]{fontenc}
\usepackage{lmodern}
\usepackage{microtype}

\usepackage{tcolorbox}
\tcbuselibrary{hooks}
\tcbset{colback=black!3!white}
\makeatletter
\tcbset{
  after app={%
    \ifx\tcb@drawcolorbox\tcb@drawcolorbox@breakable
    \else
      \@endparenv
    \fi
  }
}
\makeatother

\definecolor{c1}{RGB}{0, 0, 0}
\definecolor{c2}{RGB}{180, 0, 0}

\usepackage[left=2.3cm,right=2.3cm,top=2cm,bottom=2.5cm]{geometry}
\usepackage{booktabs}

\usepackage{enumitem}
\setlist[description]{leftmargin=1em,labelindent=1em,labelsep=0.5em}

\usepackage{xspace}
\def\etal.{et\penalty50\ al.}
\newcommand*{\eg}{e.\,g.\@\xspace}
\newcommand*{\ie}{i.\,e.\@\xspace}

\newcommand*{\rhs}{r.\,h.\,s.\@\xspace}

\usepackage[bookmarks=true]{hyperref}

\providecommand{\hypersetup}[1]{}
\providecommand{\hyperref}[2][]{#2}
\providecommand{\texorpdfstring}[2]{#1}
\providecommand{\pdfbookmark}[3][]{}
\newcommand{\email}[1]{\href{mailto:#1}{\nolinkurl{#1}}}

\hypersetup{plainpages=false}
\hypersetup{pdfpagemode=UseOutlines}
\hypersetup{bookmarksnumbered=true}
\hypersetup{bookmarksopen=true}
\hypersetup{pdfstartview=FitH}
\hypersetup{colorlinks=true}
\hypersetup{citecolor=[rgb]{0 .3 0}}
\hypersetup{urlcolor=[rgb]{.35 0 0}}
\hypersetup{linkcolor=[rgb]{0 0 .5}}
\hypersetup{citebordercolor={.6 .9 .6}}
\hypersetup{urlbordercolor={.7 .8 1}}
\hypersetup{linkbordercolor={1 .7 .7}}
\hypersetup{pdfborder={0 0 .5}}

\makeatletter
\let\@myabstract\@empty
\let\@keywords\@empty
\let\@subject\@empty
\providecommand{\affiliation}[1]{\gdef\@affiliation{#1}}
\providecommand{\myabstract}[1]{\gdef\@myabstract{#1}}
\providecommand{\keywords}[1]{\gdef\@keywords{#1}}
\providecommand{\subject}[1]{\gdef\@subject{#1}}
\def\thetitle{\@title}
\def\theauthor{\@author}
\def\theaffiliation{\@affiliation}
\def\theabstract{\@myabstract}
\def\thesubject{\@subject}
\def\thedate{\@date}
\def\thekeywords{\@keywords}
\makeatother
\AtBeginDocument{
\hypersetup{pdftitle={\thetitle}}%
\hypersetup{pdfauthor={\theauthor}}%
\hypersetup{pdfsubject={\thesubject}}%
\hypersetup{pdfkeywords={\thekeywords}}%
}

\usepackage{graphbox}

\usepackage[font=small,labelfont=bf,margin=0.05\textwidth]{caption}

\usepackage{amsmath,amssymb,mathtools}
\usepackage[mathbold,autobold,greekcaps,greeklower]{mathfixs}

\newcommand{\nn}{\nonumber}
\allowdisplaybreaks
\numberwithin{equation}{section}

\newcommand{\namedref}[2]{\hyperref[#2]{#1~\ref*{#2}}}
\newcommand{\secref}[1]{\namedref{Section}{#1}}
\newcommand{\appref}[1]{\namedref{Appendix}{#1}}

\newcommand{\figref}[1]{\namedref{Figure}{#1}}

\makeatletter
\def\mr@ignsp#1 {\ifx\:#1\@empty\else #1\expandafter\mr@ignsp\fi}%
\newcommand{\multiref}[1]{\begingroup
\xdef\mr@no@sparg{\expandafter\mr@ignsp#1 \: }%
\def\mr@comma{}%
\@for\mr@refs:=\mr@no@sparg\do{\mr@comma\def\mr@comma{,}\ref{\mr@refs}}%
\endgroup}
\makeatother
\renewcommand{\eqref}[1]{(\multiref{#1})}

\newcommand{\sfrac}[2]{{\textstyle\frac{#1}{#2}}}

\newcommand{\alg}[1]{\mathfrak{#1}}

\newcommand{\mathematica}{\textsc{Mathematica}\@\xspace}

\RequirePackage[extdef]{delimset}
\makeatletter
\providecommand{\brkleft}[1][r]{\begingroup\def\dlm@use{\delim(.}%
\if r#1 \def\dlm@use{\delim(.}\fi%
\if s#1 \def\dlm@use{\delim[.}\fi%
\if c#1 \def\dlm@use{\delim\{.}\fi%
\if a#1 \def\dlm@use{\delim<.}\fi%
\expandafter\endgroup\dlm@use}
\providecommand{\brkright}[1][r]{\begingroup\def\dlm@use{\delim.)}%
\if r#1 \def\dlm@use{\delim.)}\fi%
\if s#1 \def\dlm@use{\delim.]}\fi%
\if c#1 \def\dlm@use{\delim.\}}\fi%
\if a#1 \def\dlm@use{\delim.>}\fi%
\expandafter\endgroup\dlm@use}
\makeatother

\DeclareMathOperator{\tr}{Tr}

\DeclareMathOperator*{\Res}{Res}

\DeclareMathOperator{\sech}{sech}

\newcommand{\order}[1]{\mathcal{O}(#1)}

\newcommand{\beq}{\begin{equation}}
\newcommand{\eeq}{\end{equation}}
\newcommand{\bba}{\begin{align}}
\newcommand{\eea}{\end{align}}

\newcommand{\ii}{i}

\newcommand{\dd}[2][]{\mathinner{\mathrm{d}\ifx#1\empty\else{^#1}\fi#2}}

\newcommand{\GaudinB}{\mathbb{B}}

\title{Orthogonality of Q-Functions up to Wrapping in \texorpdfstring{\\}{}Planar
\texorpdfstring{$\mathcal{N}=4$}{N=4} Super Yang--Mills Theory}

\author{%
Till Bargheer\texorpdfstring{$^1$}{},
Carlos Bercini\texorpdfstring{$^1$}{},
Andrea Cavagli\`a\texorpdfstring{$^2$}{},
Davide Lai\texorpdfstring{$^1$}{},
Paul Ryan\texorpdfstring{$^1$}{}}

\myabstract{We construct orthogonality relations in the
Separation of Variables framework for the $\alg{sl}(2)$ sector of planar
$\mathcal{N}=4$ supersymmetric Yang--Mills theory.
Specifically, we find
simple universal measures that make Q-functions of operators with
different spins vanish at all orders in perturbation theory, prior to
wrapping corrections. To analyze this rank-one sector,
we relax some of the assumptions thus far considered in the Separation
of Variables framework. Our findings may serve as guidelines for extending this
formalism to other sectors of the theory as well as other integrable
models.}

\subject{Mathematical Physics}

\keywords{Integrability, Separation of Variables, Orthogonality,
Baxter equation, higher loops, supersymmetry, Yang-Mills theory}

\begin{document}

\pdfbookmark[1]{Title Page}{title}

\thispagestyle{empty}
\setcounter{page}{0}

\renewcommand{\thefootnote}{\fnsymbol{footnote}}
\setcounter{footnote}{0}

\mbox{}
\hfill
DESY-25-095

\vfill

\begin{center}

{\Large\textbf{\mathversion{bold}\thetitle}\par}

\vspace{1cm}

\textsc{\theauthor}

\bigskip

\begingroup
\footnotesize\itshape

${}^1$ Deutsches Elektronen-Synchrotron DESY,
Notkestr.~85, 22607 Hamburg, Germany

${}^2$ Dipartimento di Fisica, Università di Torino and INFN, Sezione
di Torino, Via P. Giuria 1, 10125, Torino, Italy

\endgroup

\bigskip

\textbf{Abstract}
\vspace{5mm}

\begin{minipage}{12cm}
\theabstract
\end{minipage}

\end{center}

\vfill
\vfill

\newpage

\renewcommand{\thefootnote}{\arabic{footnote}}

\pdfbookmark[1]{\contentsname}{contents}
\setcounter{tocdepth}{3}
\microtypesetup{protrusion=false}
\tableofcontents
\microtypesetup{protrusion=true}
\vspace{3ex}
\newpage

\section{Introduction}

The presence of integrability in planar $\mathcal{N}=4$ SYM has led to the development of powerful non-perturbative tools for studying
correlation functions. For two-point functions, the Quantum
Spectral Curve \cite{Gromov:2013pga,Gromov:2014caa} allows to extract the conformal dimension of any
single-trace local operator, at any value of the 't Hooft coupling, including strong coupling \cite{Ekhammar:2024rfj},
from a handful of so-called Q-functions. For three- and higher-point
correlation functions there is the Hexagon framework~\cite{Basso:2015zoa,Fleury:2016ykk}, valid even beyond
the planar limit \cite{Bargheer:2017nne,Bargheer:2018jvq}, and with
some modifications also in $\mathcal{N}=2$ contexts~\cite{Ferrando:2025qkr,lePlat:2025eod}.

The Hexagon framework is highly efficient for asymptotically large
operators. For short operators, contributions from so-called mirror
particles render its application much more difficult. In this sense,
the Hexagon framework is analogous to the Asymptotic Bethe Ansatz for
the spectral problem \cite{Beisert:2005fw}. For the finite-size spectrum, the latter needs to
be supplemented with Lüscher corrections via the Thermodynamic Bethe
Ansatz \cite{Bombardelli:2009ns,Arutyunov:2009ur,Gromov:2009tv}, which then leads to the Quantum Spectral Curve, with all
finite-size corrections elegantly taken into account. Additionally,
features of conformal data such as analyticity in spin \cite{Caron-Huot:2017vep} are
obscured in the Hexagon approach, while being completely manifest in the
Quantum Spectral Curve \cite{Gromov:2014bva,Klabbers:2023zdz,Brizio:2024nso,Basso:2022nny,Basso:2025mca}.

It is thus highly desirable to find a formalism that allows one to
compute correlation functions directly in terms of the Quantum
Spectral Curve Q-functions. In conventional integrable spin chains,
this is achieved by the Separation of Variables (SoV) method, which was pioneered by
Sklyanin \cite{Sklyanin:1984sb} and extensively developed in $\mathfrak{sl}(2)$-based models in \cite{Derkachov:2001yn,Derkachov:2002tf}. The formalism has undergone rapid advancement in recent
years, in particular for higher-rank symmetries~\cite{Gromov:2016itr,Maillet:2018bim,Ryan:2018fyo,Cavaglia:2019pow,Gromov:2019wmz,Gromov:2020fwh,Gromov:2022waj,Ekhammar:2023iph}.
It allows one to cast the wave functions of the Hamiltonian eigenstates
in a factorized product of single-variable building blocks, the
Q-functions, from which correlation functions can then be computed.

Separation of Variables computations have already yielded
impressive results for correlation functions in $\mathcal{N}=4$ SYM
\cite{Jiang:2015lda,Cavaglia:2018lxi,Giombi:2018qox,Caetano:2020dyp}
and related theories \cite{Cavaglia:2021mft}. The current
state-of-the-art at weak coupling is the SoV
expression for three-point correlators involving two BPS operators and
one non-BPS operator in the $\alg{su}(2)$ and $\alg{sl}(2)$ sectors \cite{Bercini:2022jxo}.

A key building
block for physical observables such as correlation functions
\cite{Gromov:2022waj} is a scalar product on the space of states
realized in terms of Q-functions,
\ie a relation
that becomes zero for
distinct states. This relation takes the form of a determinant of
inner products of Q-functions, where the inner products are integrals
against simple measures, which depend on the representation.
A modern method to construct the orthogonality relation is
the Functional Separation of Variables formalism
\cite{Cavaglia:2019pow}, which is based on the so-called
Baxter TQ equation.

In principle, since the main ingredient needed for the Functional
Separation of Variables formalism is just the TQ equation, one can
apply it to arbitrarily high orders in
perturbation theory in planar $\mathcal{N}=4$ SYM, at least in the
closed rank-one bosonic sectors.%
\footnote{The Functional Separation of Variables formalism has only
been fully developed for rational bosonic spin chains based on
$\alg{sl}(n)$ and $\alg{su}(n)$
\cite{Gromov:2020fwh,Gromov:2022waj} (see \cite{Ryan:2021duf} for a
review), and there are no closed bosonic subsectors of planar
$\mathcal{N}=4$ SYM at higher orders in perturbation theory besides
the rank-one $\alg{su}(2)$ and
$\alg{sl}(2)$ sectors.}
However, unlike in rational spin chains, where the Q-functions are
polynomial,\footnote{For
non-compact highest-weight representations, not all Q-functions are polynomial, but
all physical quantities can still be described by just the polynomial
Q-functions, at least in the examples considered in the literature,
see \cite{Gromov:2020fwh}.}
and for which most of the SoV
development has been carried out, numerous complications arise.
The Q-functions
start receiving quantum corrections and are no longer polynomial. The
Baxter TQ equations and the measures also receive corrections.
Additionally, beyond leading order in perturbation theory, we do not
have an operatorial description of integrability in $\mathcal{N}=4$
SYM, meaning that operatorial techniques such as those of
\cite{Gromov:2016itr,Maillet:2018bim,Ryan:2020rfk} are not available.
More abstractly, the fact that finite-coupling transfer matrices and
Q-operators~\cite{Derkachov:2001yn,Bazhanov:2010ts,Bazhanov:2010jq}
are not yet
formulated is related to the unknown quantum algebra underlying the
$\mathcal{N}=4$ SYM model.
A full understanding of the underlying
quantum algebra should indicate an algebraic path towards the corresponding
Q-operators.%
\footnote{There has been some recent progress towards the related quantum algebra of the
spin chain excitation picture~\cite{Beisert:2024rke}.}

In view of these difficulties, in this paper, we take an explorative
approach to construct the desired orthogonality relation. We focus on
the $\alg{sl}(2)$ sector, whose operators are described by their twist
$L$ and spin $S$ and take the schematic form
\begin{equation}
    \tr(D^S Z^L)+\text{permutations}
\end{equation}
where $Z$ is a complex scalar and $D$ is a covariant derivative.
Based on observations for twist-two and twist-three states at low loop
orders, we propose an orthogonality relation for all $\alg{sl}(2)$
states to all orders in perturbation theory before wrapping effects
kick in. The orthogonality is formulated in terms
of finite-coupling measures~\eqref{finiteMu} together with a set of
ever-growing matrices~\eqref{enlargedAll} of Q-functions whose determinant~\eqref{claim} vanishes when the states have different values of spin.

However, there is one important aspect in our SoV-type
expressions that
needs to be emphasized. For reasons we do not yet fully understand,
our proposed orthogonality~\eqref{claim} breaks for states
of equal spin. Since in $\alg{sl}(2)$, the spin is a proxy for
leading-order degeneracy between the states, another way we can state this fact is:
\emph{Our SoV expressions are valid only for operators that are not
degenerate at leading order}. For twist-two operators, there are no such
degeneracies, so this problem is absent. On the contrary,
higher-twist operators do present this problem. For example, at twist three,
there are three degenerate states
with spin six at leading order.
We observe that our formulas correctly reproduce orthogonality
between these degenerate states and all the other twist-three
operators (up to wrapping), but they give a non-vanishing overlap when contracting the
degenerate states among themselves.

In the last section of this work, we present a possible solution to
this puzzle, by considering generalizations to the usual assumptions
considered in the SoV framework. This allows us to establish a new
strategy to deduce orthogonality in terms of Q-functions that we believe
could also be applicable in more general integrable systems, where
the Separation of Variables approach has proven to be elusive.

In~\secref{secTwist2} we review the SoV approach for twist-two
operators, recalling the results for two-point functions at leading
and next-to-leading order. We show how the problems involving the
appearance of new degrees of freedom naturally emerge at
N$^2$LO, and how we solve them.

In~\secref{secOrthoAll} we present the main results of this work,
where we generalize the expressions of twist-two to higher-twist
operators at any order in perturbation theory before wrapping
corrections.

In~\secref{WayOut} we generalize the usual assumptions of the SoV
framework, and consider a more general type of Q-function for twist-two
operators. This results in a new SoV expression at
N$^2$LO that we believe can resolve the puzzle of
orthogonality among degenerate states and yield new SoV
representations for integrable models.

Finally, in~\secref{Discussion} we summarize all our results and
comment on the several limits of strong coupling, large spin and large
charge that could be accessible by our SoV proposal.

\section{Orthogonality for Twist Two}
\label{secTwist2}

\subsection{Functional Separation of Variables at Leading Order}
We will start by considering the simplest class of $\alg{sl}(2)$ operators,
the twist-two operators, which have the form $\text{Tr}(Z D^S Z) +
\texttt{permutations}$. At leading order, the Q-functions that
describe such operators are given by the solutions to the Baxter
equation~\cite{Basso:2011rc}
\begin{equation}
  0=\mathcal{B}_S\cdot Q_S \equiv \left(u+\frac{i}{2}\right)^2 Q_S(u+i)+\left(u-\frac{i}{2}\right)^2 Q_S(u-i)+(I_0(S) - 2u^2)Q(u)\,.
  \label{baxterEqLead}
\end{equation}
Here, we have introduced a
finite-difference operator $\mathcal{B}_S$ called the \emph{Baxter
operator}, which acts on the Q-function by shifts of the spectral
parameter. As we review below, $\mathcal{B}_S$ is a central object
for the functional SoV method.
The term that multiplies $Q(u)$ on the right-hand side is the
\emph{transfer matrix}, whose expansion coefficients are the
state-dependent integrals of motion $I_j(S)$. In this
example, $I_0(S)=1/2+S+S^2$ and $I_2(S)=-2$.

The solution space of the Baxter equation~\eqref{baxterEqLead} is
two-dimensional. A generic solution will be non-polynomial, but there
is one unique polynomial solution
\begin{equation}
  Q_S(u) = \prod_{n=1}^{S}(u-v_n)\,,
  \label{leadQ}
\end{equation}
whose roots $v_n$ are the solution to the asymptotic Bethe equations
recalled in~\appref{appBaxter}. We will come back to the other,
non-polynomial solutions in~\secref{WayOut}.

Our goal is to find Separation of Variables type expressions for
two-point functions in this $\alg{sl}(2)$ sector of $\mathcal{N}=4$ super
Yang--Mills theory. Namely, we want to find simple universal measures that, when
integrated against these Q-functions, result in the
two-point correlation function of the corresponding operators. Since two-point functions of different
states are zero, one simpler and often sought-after goal is to find
measures that make Q-functions of different states \emph{orthogonal} to each
other. In the Functional Separation of Variables approach, the
starting point for this orthogonality is the
trivially vanishing expression
\begin{equation}
  \int_{\mathbb{R}} \left[Q_S(u)\overleftarrow{\mathcal{B}}_SQ_J(u)-Q_S(u)\overrightarrow{\mathcal{B}}_JQ_J(u)\right] \mu(u) du = 0 ,
\label{leadingSelfAdj}
\end{equation}
where $\overleftarrow{\mathcal{B}}_S$
($\overrightarrow{\mathcal{B}}_J$) act on the Q-function on their left
(right), respectively, and $\mu(u)$ is an integration measure which
will be fixed by imposing some essential properties. In fact, the trivial
relation (\ref{leadingSelfAdj}) can be turned into a useful relation
if we can find an integration measure that is both state-independent
and makes the Baxter operator self-adjoint,
\ie allows us to
replace $\overleftarrow{\mathcal{B}}_S$ with
$\overrightarrow{\mathcal{B}}_S$ under integration.
This can be
achieved by finding an $i$-periodic measure that allows us to freely
shift contours in the imaginary direction, in order to move the shifts in $u$ from one Q-function
to the other. For example,
\begin{equation}
  \int_{\mathbb{R}} Q_S(u)\left(u+\frac{i}{2}\right)^2 Q_J(u+i)\mu(u) du  \overset{u\to u-i}{=} \int_{\mathbb{R}} Q_S(u-i)\left(u-\frac{i}{2}\right)^2 Q_J(u)\mu(u) du\,.
  \label{shift}
\end{equation}
This shifting of contours has been extensively considered in the SoV
framework in many different
contexts~\cite{Gromov:2019wmz,Cavaglia:2019pow,Gromov:2020fwh,Levkovich-Maslyuk:2025ipl},
and its outcome is to recast
equation~\eqref{leadingSelfAdj} as
\begin{equation}
  0=\int_{\mathbb{R}} Q_S(u)(\overleftarrow{\mathcal{B}}_S-\overleftarrow{\mathcal{B}}_J)Q_J(u)\mu(u) du + \texttt{res}_\mu = \Delta I_0 \int_{\mathbb{R}} Q_S(u)Q_J(u)\mu(u) du + \texttt{res}_\mu
  \,,
  \label{leadingSelfAdj2}
\end{equation}
where we used the explicit expression of the leading-order Baxter
operator~\eqref{baxterEqLead} to write the result in terms of a
difference of integrals of motion $\Delta I_0\equiv I_0(S)-I_0(J)$. The
additional term $\texttt{res}_{\mu}$ denotes potential residues that in general
will be picked up when shifting the contours in the above maneuvers. In
all applications of the Functional SoV method in the existing literature, such residues can
be canceled by a judicious choice of $\mu(u)$, which restricts the measure from
a generic periodic function to a discrete set of possibilities.

In the simple example of the $\alg{sl}(2)$ spin chain, the restriction
of the measure $\mu(u)$ is achieved by
requiring that it exponentially decays at infinity (such that the
integral against any of the polynomial Q-functions converges), and has
at most double poles at $u=\pm \ii/2$. Such poles are canceled by the double zeros
$(u\pm i/2)^2$ of the Baxter operator~\eqref{baxterEqLead} in the
potential residue-generating terms, so that the full integrand is free
of poles in the region affected by the shifts. The unique
measure satisfying these properties is
$\mu(u)={\pi}/{2}\,\sech(\pi u)^2$, which was also used
in~\cite{Derkachov:2002tf,Jiang:2016ulr,Cavaglia:2019pow,Bercini:2022jxo}.%
\footnote{The most general $i$-periodic measure that decays
exponentially at infinity is $\mu(u)=\tanh^m(\pi u)\sech^{2n}(\pi u)$ with
$m\geq0$, $n\geq1$. It has a pole of degree $m+2n$ at $u=\pm i/2$,
hence we require $m+2n\leq2$, which leaves $n=1$, $m=0$ as the only choice.}
The relation~\eqref{leadingSelfAdj2} then becomes
\begin{equation}
  \Delta I_0 \int_{\mathbb{R}} Q_S(u)Q_J(u)\mu(u) du = 0 .
  \label{zeroFirst}
\end{equation}
Since distinct states have distinct integrals of motion (and
conversely, each state is completely specified by its
integrals of motions), the integral above must vanish when the two
states are different, \ie we obtain the desired orthogonality relation
between Q-functions:
\begin{equation}
  \langle Q_S Q_J \rangle_\mu \equiv \int_{\mathbb{R}} Q_S(u)Q_J(u)\mu(u) du \propto \delta_{S,J}\,.
  \label{orthoLeading}
\end{equation}

Our goal is to obtain a similar orthogonality relation for Q-functions
at higher orders in weak-coupling perturbation theory. We focus our
attention on the next-to-next-to-leading order,\footnote{We will refrain to
use the term ``loops'', due to the confusion it may cause between the
order of corrections in the Q-function and the actual loop order in
the field theory, \eg order $O(g^4)$ in the Q-function computes
corrections of order $O(g^6)$ in the spectrum of the gauge theory.} order
since it is the first order where almost no results are known for
orthogonality,\footnote{There is a N$^2$LO orthogonality
for twist-2 operators written in~\cite{Bercini:2022jxo}. However, the
measure used there is not universal but rather depends on the states
appearing in the inner product.} while it still avoids wrapping corrections
for the operators. As we will mention below, the
game becomes significantly more complicated in the presence of wrapping. Most importantly, the
lessons from our N$^2$LO explorations will
serve as guidelines to extend orthogonality to both higher twist and higher
orders in perturbation theory.

\subsection{Higher Loops and Asymptotic Baxter Equation}

Beyond the leading order in perturbation theory, the polynomial
Q-function above is not sufficient to characterize the spectrum.
Instead, the finite-coupling spectrum of conformal primaries in planar
$\mathcal{N}=4$ SYM is encoded in the Quantum Spectral Curve \cite{Gromov:2013pga,Gromov:2014caa}. This is
a set of 256 Q-functions $Q_{A|I}(u)$, $A,I\subset\{1,2,3,4\}$, anti-symmetric in both $A$ and $I$, linked by various algebraic and analytic relations. In the asymptotic
large-volume regime that we restrict to in this work, a central role
is played by the Q-function $Q_{12|12}(u)$. This
Q-function encodes the momentum-carrying roots $v_n$, and at leading
order is closely related to the polynomial Q-function~\eqref{leadQ} that we considered
before. In this regime, it takes the form
\begin{equation}
    Q_{12|12}(u) = \mathcal{P}(u) (f^+)^2\,,\quad \text{with}\quad f^\pm:= f(u\pm\tfrac{i}{2})\, ,
    \label{Q1212}
\end{equation}
where $\mathcal{P}$ is the polynomial part of the Q-function
\begin{equation}
    \mathcal{P}(u) = \prod_{k=1}^S(u-v_k)
    \,.
    \label{polyQ}
\end{equation}
The roots $v_k$ are the momentum-carrying roots (which acquire loop corrections in
perturbation theory) of the Asymptotic Bethe Ansatz (ABA) equations
associated to the middle node of the $\alg{su}(2,2|4)$ Dynkin diagram. In the
case of the $\alg{sl}(2)$ sector we consider here, the ABA equations are
recalled in~\appref{appBaxter}.
Additionally, we have the zero-momentum condition
$\mathcal{P}(-{i}/{2})=\mathcal{P}({i}/{2})$.

The function $f(u)$ is a non-polynomial dressing, with no poles or
zeros anywhere, and no branch cuts in the upper half plane
\cite{Gromov:2014caa}, that satisfies
\begin{equation}
    \frac{f(u+i)}{f(u)}= \prod_{k=1}^S\left(\frac{\frac{1}{x(u)} - x_k^+ }{\frac{1}{x(u)} - x_k^- }\right) \left(\frac{x_k^-}{x_k^+}\right)^{\frac{1}{2}}\,,
    \label{dressing}
\end{equation}
where $x(u)$ is the Zhukovsky map (see~\appref{appBaxter}). A
solution to this equation, with power-like asymptotics at infinity and
analytic in the upper half plane, is given by
\begin{equation}
    f(u) = {\rm exp}\left( g^2 q_1^+ \psi_0(-iu)+\frac{g^4}{2}\left(i q_2^-\psi_1(-iu)-q_1^+ \psi_2(-iu)\right) +\mathcal{O}(g^6)\right)
\end{equation}
where $\psi_k$ are the polygamma functions of order $k$ and $q_k^\pm$
are conserved charges, also recalled in~\appref{appBaxter}.

To quick-start our SoV explorations, we will consider a slight
generalization of the momentum-carrying Q-function, where we
symmetrize its dressing, and rescale it by a parameter:
\begin{equation}
  Q_S(u) = \prod_{n=1}^{S}(u-v_n)e^{\alpha\cdot\sigma(u)}
  \,,\qquad
  \sigma(u)=\log(f^+ \bar{f}^-)
  \,,
  \label{QFunction}
\end{equation}
where the symmetrized dressing $\sigma(u)$ can be easily computed
from~\eqref{dressing} at any order in perturbation theory. For example,
the first few orders are
\begin{align}
  \sigma(u) & = 2g^2 q_1^{+}\psi_{0,+}(u)-g^4\left(q_1^{+}\psi_{2,+}(u)+q_2^{-}\psi_{1,-}(u)\right)+\mathcal{O}(g^6)
  \,,
  \label{dressingExpand}
\end{align}
where the symmetric and anti-symmetric
combinations of polygamma functions are
\begin{align}
    \psi_{n,\pm}(u) &= \frac{1}{2}\left(\psi_n\left(\frac{1}{2}+ i u\right)\pm \psi_n\left(\frac{1}{2}- i u\right)\right)\,.
\end{align}

The QSC Q-function $Q_{12|12}(u)$ is analytic in the upper half
plane (\ie its dressing $f^+$ has poles only on the negative imaginary
axis, spaced by $i/2$), while the Q-function~\eqref{QFunction} with symmetrized dressing
$\sigma$ has poles for both negative and positive imaginary values.
Since $f^+/\bar{f}^-$ is a periodic function, when the dressing
parameter $\alpha$ is one, the QSC Q-function~\eqref{Q1212} and the
symmetrized Q-function~\eqref{QFunction} satisfy the same Baxter
equation. For generic values of dressing parameter $\alpha$, this is no
longer the case, but it is still easy to write the Baxter equation
that the Q-functions~\eqref{QFunction} satisfy:
\begin{equation}
  0 = \mathcal{B}_S\cdot Q_S = B^{+}_SQ_S(u+i)+B^{-}_SQ_S(u+i)+T_S(u)Q_S(u)\,,
  \label{B2loop}
\end{equation}
where the shift operators are given by%
\footnote{In principle, we could also refer to the conserved charges as
integrals of motion, since they are ultimately functions of the
state's quantum numbers. However, we will make a distinction between
these objects, by denoting as integrals of motion the quantities that
appear in the transfer matrix ($T_S$) and as charges the quantities
that appear in the shift operators ($B^\pm_S$).}
\begin{equation}
  B^{\pm}_S = \brk{x^{\pm}}^2 \pm i g^2 q_1^{+}(1-2\alpha)\brk{x^{\pm}}^1
  +g^4\brk1{q_2^{+}+(q_1^{+})^2(1-2\alpha)^2}\brk{x^{\pm}}^0
  \,,\label{shiftOp}
\end{equation}
and the transfer matrix is
\begin{equation}
  T_S =
  I_0(S) - 2u^2
  +g^4\brk*{\frac{1}{\left(x^{-}\right)^2}+\frac{1}{\left(x^{+}\right)^2}}
  \,.
  \label{Transfer2Loop}
\end{equation}
Since three different values of $\alpha$ will be important in the following,
let us establish some notation before diving into the SoV
explorations. Starting from~\eqref{QFunction}, we define three
Q-functions that are the same at leading order, but differ in
perturbation theory:
\begin{equation}
    \mathcal{P}_S(u) = \left.Q_S(u)\right|_{\alpha=0}
    \,, \quad
    \mathcal{Q}_S(u) = \left.Q_S(u)\right|_{\alpha=1}
    \,, \quad
    \mathbb{Q}_S(u) = \left.Q_S(u)\right|_{\alpha=\frac{1}{2}}
    \,.
\end{equation}
As mentioned before, when $\alpha=1$, this is the usual (symmetrized)
momentum-carrying QSC Q-function. When $\alpha=0$, the dressing
vanishes and we are left only with the polynomial part of this
Q-function. The Q-function with $\alpha=1/2$, also considered
in~\cite{Bercini:2022jxo}, will be important below and
in~\secref{WayOut}, but as yet has no realization in terms of
combinations of QSC objects.

In the presence of wrapping, several complications arise.
First of all, the form of the Q-functions will become complicated, and
will no longer be parametrized by Bethe roots. In principle, one can
still compute the Q-functions explicitly at any order at weak
coupling, using the package of \cite{Marboe:2014gma}. However, since
we lose the meaning of Bethe roots beyond wrapping order, it is far
less clear why a single Q-function should play a prominent role in the
SoV expression. It is more natural to expect that the full result
involves the entire $\alg{psu}(2,2|4)$ structure, which is far from
being understood in functional SoV methods. For these reasons, we
focus our attention on the situation where no wrapping effects are
present.

\subsection{Setting up Higher Loop Orthogonality}

By simply inspecting the shift operator~\eqref{shiftOp}, one might be
tempted to consider $\alpha={1}/{2}$, as was done
before~\cite{Bercini:2022jxo},
and in fact, that is exactly what we will
initially do. Although it \emph{will not} lead to
N$^2$LO orthogonality, it will illustrate the problems
one must overcome when considering orthogonality at higher orders in
perturbation theory. Simply setting $\alpha={1}/{2}$
in~\eqref{B2loop} results in the simple Baxter equation
\begin{multline}
\left(\brk{x^{+}}^2 + g^4\frac{q_2^{+}}{2}\right)\mathbb{Q}_S(u+i)+\left(\brk{x^{-}}^2 + g^4\frac{q_2^{+}}{2}\right)\mathbb{Q}_S(u-i)=\\
-\left(I_0(S) - 2u^2 + g^4\left(\frac{1}{\left(x^{-}\right)^2}+\frac{1}{\left(x^{+}\right)^2}\right)\right)\mathbb{Q}_S(u)\,.
\end{multline}

Note also that the QSC-inspired dressing~\eqref{QFunction} for any
value of $\alpha$ has a
property that makes the SoV explorations remarkably simple: The
non-polynomial part of the transfer matrix is state-independent, while
the shift terms remain simple polynomials in $x^\pm$ and conserved
charges.
Therefore, when we consider differences of Baxter operators as in
equation~\eqref{leadingSelfAdj}, these non-polynomial
parts will cancel out (just like the universal term $-2u^2$ canceled at
leading order). This reduces the number of integrals of motions that
appear in the SoV expressions and, as we will see, simplifies the
search for orthogonal measures.

Starting from the initial equation~\eqref{leadingSelfAdj}, and
performing the same contour manipulations as we did at leading order,
we obtain the following N$^2$LO expression:
\begin{equation}
  \Delta I_0 \int_{\mathbb{R}} \mathbb{Q}_S(u)\mathbb{Q}_J(u)\mu(u) du + g^4\frac{\Delta q_2^{+}}{2} \int_{\mathbb{R}} (B_0 \cdot \mathbb{Q}_S) \mathbb{Q}_J(u)\mu(u) du + \texttt{res}_\mu = 0\,,
  \label{twoLoopSelfAdj}
\end{equation}
where $B_0$ is a particular case of what will be an important new
ingredient of orthogonality beyond leading order: The
\emph{lower-length Baxter operators}, defined as
\begin{equation}
  B_M \cdot F = \left(u+\frac{i}{2}\right)^M F(u+i)+\left(u-\frac{i}{2}\right)^M F(u-i)-2u^M F(u)\,,
  \label{lowerBaxter}
\end{equation}
whose ``transfer matrix'' (coefficient of the $F(u)$ term on the
r.h.s.)
differs from the usual Baxter
operator~\eqref{baxterEqLead}, since it depends \emph{only} on the
universal $u^M$ term. We will comment more on this difference, and the
central role of these operators once we consider orthogonality for
general twists in~\secref{secOrthoAll}.

In general, to obtain orthogonality relations akin
to~\eqref{orthoLeading}, one needs to find as many measures as
independent integrals of motion and charges appear in the
equation. For example, in the
expression~\eqref{twoLoopSelfAdj} we have \emph{one} integral of
motion $\Delta I_0$ and \emph{one} conserved charge $\Delta q_2^{+}$
(notice that we have eliminated $q_1^+$ and $(q_1^+)^2$ from the
equation by our choice of $\alpha = {1}/{2}$), so \emph{two}
measures with vanishing residues are needed to define orthogonality.
Let's assume that two such measures $\mu_1(u)$ and $\mu_2(u)$ exist
(each intended as an expansion in the coupling up to $O(g^4)$). Then
we could write the following linear system
\begin{equation}
  \left\{ \begin{aligned}
    \Delta I_0 \langle \mathbb{Q}_S \mathbb{Q}_J \rangle_{\mu_{1}} + g^4 \frac{\Delta q_2^+}{2} \langle (B_0 \cdot \mathbb{Q}_S)\mathbb{Q}_J \rangle_{\mu_{1}} &= 0\\
    \Delta I_0 \langle \mathbb{Q}_S \mathbb{Q}_J \rangle_{\mu_{2}} + g^4 \frac{\Delta q_2^+}{2} \langle (B_0 \cdot \mathbb{Q}_S)\mathbb{Q}_J \rangle_{\mu_{2}} &= 0
  \end{aligned} \right.
  \label{linearSystem}
\end{equation}
where we used the notation introduced in~\eqref{orthoLeading} to
denote the integrals of the Q-functions over the real line. It will be
useful to rewrite the linear system above as the following matrix
equation
\begin{equation}
  \begin{pmatrix}
    \langle \mathbb{Q}_S \mathbb{Q}_J \rangle_{\mu_{1}} & \langle (B_0 \cdot \mathbb{Q}_S)\mathbb{Q}_J \rangle_{\mu_{1}} \\
    \langle \mathbb{Q}_S \mathbb{Q}_J \rangle_{\mu_{2}} & \langle (B_0 \cdot \mathbb{Q}_S)\mathbb{Q}_J \rangle_{\mu_{2}}
  \end{pmatrix}
  \cdot
  \begin{pmatrix}
    \Delta I_0 \\
    {g^4/4\,\Delta q_2^{+}}
  \end{pmatrix}
  =0\,.
  \label{possibility}
\end{equation}
Since distinct states have non-coincident vectors of integrals of
motion, the fact that the homogeneous linear system has a solution implies that the
determinant of the matrix of inner products is zero, \ie it implies
the following orthogonality relation for the Q-functions:
\begin{equation}
  \begin{vmatrix}
    \langle \mathbb{Q}_S \mathbb{Q}_J \rangle_{\mu_{1}} & \langle (B_0 \cdot \mathbb{Q}_S)\mathbb{Q}_J \rangle_{\mu_{1}} \\
   \langle \mathbb{Q}_S \mathbb{Q}_J \rangle_{\mu_{2}} & \langle (B_0 \cdot \mathbb{Q}_S)\mathbb{Q}_J \rangle_{\mu_{2}}
  \end{vmatrix}
  \propto \delta_{S,J} +\mathcal{O}(g^6) \,.
  \label{possibilityDet}
\end{equation}
To find the measures with vanishing residues, we follow the method
described in~\appref{appResidues}: We write the ansatz for the measure as a Laurent
series, and computes the residues as an explicit combination of the
coefficients of this series expansion and the conserved charges of the
Q-functions. Since the conserved charges are all
independent,\footnote{For a fixed twist, all conserved charges are
linearly independent functions of the spin. For example, for twist-two
operators at leading order, we have: $q_1^{+}= 4 H_1(S)$ and
$q_2^{+}=2\brk2{H_2\left(\frac{s-1}{2}\right)-H_2\left(\frac{s}{2}\right)+\frac{\pi^2}{3}}
$, where $H_n(S)$ are the harmonic numbers.} demanding that the
residues cancel can be done analytically by requiring that the
coefficient of each conserved charge vanishes. This process fixes all
the coefficients of the Laurent series of the measure, which in turn
determines the sought-after measures with vanishing
residues.

It turns out that, for the particularly simple case we have been
considering so far (Q-functions with dressing parameter
$\alpha={1}/{2}$), and in fact for any value of dressing parameter~$\alpha$,
this analytic residue computation shows that only
\emph{one} measures with vanishing residues exists (which we describe in
more details in~\secref{WayOut}). Since we do not have as many
measures with vanishing residues as integrals of motion and conserved
charges, it seems that N$^2$LO orthogonality for twist-two
operators, if it exists, must have different ingredients than its
previous orders in perturbation theory.

\subsection{Exploring Twist-Three Orthogonality}

To better understand what types of structures we can expect at higher
loops, let us consider twist-three operators at one loop.

The main new technical obstacle that arises for twist three is that,
unlike at twist two, not all states are parity-symmetric. Non-symmetric
states start to appear at spins $S\geq 6$ and these states have a
non-zero charge $q_1^-$. This charge enters the coefficients of the
Baxter operator at one loop
\begin{equation}
    \mathcal{B}\cdot \mathbb{Q} = (x^+)^3\left(1+g^2\frac{q_1^{-}}{x^+}\right)\mathbb{Q}(u+i) +(x^-)^3\left(1+g^2\frac{q_1^{-}}{x^-}\right)\mathbb{Q}(u-i) - T(u)\mathbb{Q}(u) = 0
\end{equation}
which annihilates the dressed Q-function
$\mathbb{Q}(u)$ with $\alpha=1/2$ considered in the
previous section. The transfer matrix is given by
\begin{equation}
    T(u)=2u^3 -2g^2 q_1^- +I_1 u+ I_0\,,
\end{equation}
and we remind that $I_1$ and $I_0$ are functions that depend both on the state and on the coupling.

Note that from a practical perspective, the obstacles posed by the
presence of $q_1^-$
at one loop are very similar to those of $q_2^+$ at two loops for
twist-two, as both enter the Baxter equation in an almost identical
fashion.

Since $q_1^-$ always appears with a factor of $g^2$, it
does not contribute at leading order. Then we only need to find two independent
measures such that the residues vanish. Following the logic for
twist-two, these measures can be computed to be given by
\begin{equation}
    \mu_1(u) =  \frac{\pi}{2\cosh^2(\pi u)}, \quad \mu_2(u)  = \frac{\pi^2 \tanh(\pi u)}{\cosh^2(\pi u)}\,.
\end{equation}
This results in the leading-order orthogonality relation
\begin{equation}
    \begin{pmatrix}
        \langle \mathcal{P}_S \mathcal{P}_J \rangle_{\mu_1} & \langle \mathcal{P}_S \,u  \,\mathcal{P}_J \rangle_{\mu_1} \\
        \langle \mathcal{P}_S \mathcal{P}_J \rangle_{\mu_2} & \langle \mathcal{P}_S \,u  \,\mathcal{P}_J \rangle_{\mu_2}
    \end{pmatrix} \cdot
    \begin{pmatrix}
        \Delta I_0 \\
        \Delta I_1
    \end{pmatrix} = 0\,.
\end{equation}

We now proceed exactly as we did for twist two, and try to find
measures such that the residues cancel. In this case, we need to find
three measures $\mu_i$, since we have three non-trivial integrals of
motion in the Baxter equation at this loop order: Two integrals of
motion $I_0$ and $I_1$ coming from the transfer matrix, as well as
$q_1^-$. At leading order, we can indeed find three such measures $\mu_\ell$, $\ell=1,2,3$,
\emph{provided} we restrict our analysis to zero-momentum states. These measures are given explicitly by
\begin{equation}
    \mu_\ell(u) = \frac{\ell \pi^{\ell}/2}{\cosh^2(\pi u)}\tanh^{\ell-1}(\pi u)
    \,,
    \label{twist3mu}
\end{equation}
where the overall normalization is rather unimportant. Indeed, for
$\mu_3$ the residue contribution at leading order is
proportional to
\begin{equation}
    \frac{\mathcal{P}_S(-{i}/{2})}{\mathcal{P}_S({i}/{2})}-\frac{\mathcal{P}_J(-{i}/{2})}{\mathcal{P}_J({i}/{2})}
    \,,
\end{equation}
and so it vanishes when both states satisfy the zero-momentum condition
$\mathcal{P}(-{i}/{2})=\mathcal{P}({i}/{2})$.

Proceeding to the first order in the weak-coupling expansion, we find
that by considering simple perturbative corrections to the
measures~\eqref{twist3mu}, given by
\begin{equation}
    \mu_\ell(u) = \frac{\ell \pi^{\ell}/2}{\cosh^2(\pi u)}\tanh^{\ell-1}(\pi u)\left(1+g^2\pi^2\left(-\frac{\ell}{3}+(\ell+2)\tanh^2(\pi u)\right)\right)+ \mathcal{O}(g^4)\,,
    \label{mutwist3loop}
\end{equation}
we can make the residues vanish at NLO, provided again that the states
satisfy the zero-momentum condition. Hence, we obtain the following
linear system:
\begin{equation}
  \begin{pmatrix}
    \langle  \mathbb{Q}_S  \mathbb{Q}_J \rangle_{\mu_{1}} & \langle  \mathbb{Q}_S\, u\,  \mathbb{Q}_J \rangle_{\mu_{1}} & \langle (B_2 \cdot  \mathbb{Q}_S)  \mathbb{Q}_J \rangle_{\mu_{1}} \\
    \langle  \mathbb{Q}_S  \mathbb{Q}_J \rangle_{\mu_{2}} & \langle  \mathbb{Q}_S\, u\,  \mathbb{Q}_J \rangle_{\mu_{2}} & \langle (B_2 \cdot  \mathbb{Q}_S)  \mathbb{Q}_J \rangle_{\mu_{2}} \\
    \langle  \mathbb{Q}_S  \mathbb{Q}_J \rangle_{\mu_{3}} & \langle  \mathbb{Q}_S \, u\,  \mathbb{Q}_J \rangle_{\mu_{3}} & \langle (B_2 \cdot  \mathbb{Q}_S)  \mathbb{Q}_J \rangle_{\mu_{3}}
  \end{pmatrix}\begin{pmatrix}
      \Delta I_0 \\
      \Delta I_1 \\
      g^2 \Delta q_1^-
  \end{pmatrix}=0
  \label{eq:twistThreeMatrix}
\end{equation}
where $B_2$ is the ``lower-length'' Baxter operator of~\eqref{lowerBaxter}:
\begin{equation}
    B_2\cdot F = \left(u+\frac{i}{2}\right)^2 F(u+i)+\left(u-\frac{i}{2}\right)^2 F(u-i)-2u^2 F(u)
\end{equation}
as usual, the
determinant of the coefficient matrix must vanish for two distinct
states, and thus provides an NLO orthogonality relation for
twist-three states. This construction can be extended to any twist
$L$, and takes the form of an $L\times L$ determinant. Interestingly, this seems to provide a different one-loop
orthogonality relation than found in \cite{Bercini:2022jxo}, which
takes the form of a $(L-1)\times(L-1)$ matrix. The relation between the two
is not completely clear, as the relation found in that paper was not
systematically derived.

Unfortunately, we were not able to extend this observation to higher loops. The main
issue we encounter, which is a general obstacle not restricted
to the present set-up, is that at two loops, more
integrals of motion will appear in the equation, which necessitates more measures. Any additional
measure we introduce does not have vanishing residues, even at tree
level, and these residues are proportional to yet more integrals of
motion, necessitating even more measures, and so on. In short, there
is an uncontrolled runaway of integrals of motion, even at fixed
orders of perturbation theory.

Let us note that the observations above may be a hint that for higher
loops we need to consider more general solutions of the Baxter
equation. Indeed, the construction above crucially relies on the multiplication of the powerlike solution by the periodic function defined above \eqref{B2loop}, and the residues in the system of equations does not vanish without it. In principle, we can also involve the second
(non-polynomial) solution of the Baxter equation, and we will do
exactly that in \secref{sec:Q2}. While this Q-function may seem
unnatural at weak coupling, at finite coupling, all Q-functions are
essentially on equal footing. Understanding exactly which solutions of
the Baxter equation are needed for orthogonality at higher loops may
provide valuable insights into a finite-coupling SoV construction.

Before describing explorations in this direction in
\secref{sec:Q2}, we consider other possible ways to widen the usual
methodology of functional SoV.
We will use some of these ideas, abandoning some of the usual
assumptions, to directly try to construct bilinear integrals of
Q-functions with vanishing determinant for two distinct states, which
is the end goal of the construction anyway.

\subsection{What Could be Done Differently?}

Even though we encounter the roadblock of not finding enough measures
with vanishing residues at N$^2$LO (both at twist two and at
twist three), there are still several alternatives to be considered in the
Separation of Variables formalism.

One possibility is to radically change the approach, and consider a
different integration contour, such as a vertical contour parallel to
the imaginary axis, as was employed to obtain SoV-like
expressions for certain two-point and three-point functions in the context
of cusped Wilson loops in the ladders limit~\cite{Cavaglia:2018lxi}, or
in the fishnet theory~\cite{Cavaglia:2020hdb,Cavaglia:2021mft}. The
main advantage of such contours
is that in all manipulations needed to prove orthogonality, the contour
shifts onto itself, so there would be no residues to pick. However, to
guarantee convergence with such integration contours, it
is necessary to deform the system by twisting the boundary conditions
of the Q-functions, which modifies their
asymptotics~\cite{Gromov:2020fwh,Cavaglia:2019pow,Cavaglia:2020hdb}.
 In this work, we will not twist the Q-functions, and thus we will not
follow this route. Instead, we will push the results of integration
along the horizontal line beyond leading-order perturbation theory.

The way to do that is to embrace the fact that the residues will not
be zero anymore. This will shift our focus from demanding that
measures have vanishing residues to demanding that the residues have a
very specific form. This will allow us to construct linear systems
akin to~\eqref{possibility}, and will result in orthogonality
relations among the Q-functions. To obtain such orthogonality
relations, we will follow two distinct approaches, one more analytic
in nature, while the other is based on numerical explorations.

The more analytic approach is to enlarge the system by considering the
discarded second solution to the Baxter equation. The motivation
behind this is to treat the rank-one $\alg{sl}(2)$ sector in the same way as the
higher-rank
sectors~\cite{Gromov:2020fwh,Cavaglia:2019pow,Levkovich-Maslyuk:2025ipl},
whose orthogonality is given by a combination of all the different
types of Q-functions of the system. Considering both Q-functions of
the $\alg{sl}(2)$ sector, together with allowing for non-vanishing residues, will result in
the N$^2$LO orthogonality for twist-two operators
presented in~\secref{WayOut}. However, this approach becomes rapidly very
complicated,
hence we were not able to
generalize it to operators with arbitrary twist at higher orders
in perturbation theory.

Another more experimental approach we took is to forget the different types of
Q-functions, the initial equation~\eqref{baxterEqLead} and its
residues, and just search
numerically for measures and enlarged matrices
of Q-functions, like the one proposed in~\eqref{possibilityDet}, that
give orthogonality
beyond leading order. To construct a sensible ansatz for these
putative enlarged matrices, we will draw inspiration from the
twist-three case at NLO discussed above, where
we have complete analytic control, and where we have indeed found such
a matrix equation~\eqref{eq:twistThreeMatrix}.
Building upon the enlarged matrices we encounter in this setup, we
will be able to define orthogonality not only for twist-two operators at
N$^2$LO, but also for operators of arbitrary twist at
any order in perturbation theory, up to wrapping order (with the
aforementioned drawback about degenerate states). This result
will also reveal the interesting fact that the dressing parameter
$\alpha$ needs to appear asymmetrically for the two states. This is
what we will turn to now.

\subsection{Higher-Loop Measures}

Our strategy is to construct putative enlarged matrices similar
to~\eqref{possibility} and~\eqref{eq:twistThreeMatrix} for twist-two
operators, and then search
for measures that make their determinant orthogonal at
N$^2$LO. To start these explorations, we must first
establish how these matrices are constructed. They will obey three
simple rules:

\begin{description}
\item[Rule 1.]

All Q-functions are of the
form~\eqref{QFunction}. This fixes the functional form of the dressing
to be~\eqref{dressing}, while allowing each Q-function to have its own value of
dressing parameter~$\alpha$. For any value of~$\alpha$, the non-polynomial part of the transfer
matrix is universal, as in~\eqref{Transfer2Loop}. Such
state-independent terms cancel in the differences of the initial
equation~\eqref{baxterEqLead}, and thus minimize the number of
integrals of motion that need to be considered.

\item[Rule 2.]

The matrix elements can be either Q-functions
$\langle Q_S Q_J \rangle$, or Q-functions modified by lower-length Baxter operators $\langle (B_M
\cdot Q_S)Q_J \rangle$~\eqref{lowerBaxter}, as motivated from the twist-three
expression~\eqref{eq:twistThreeMatrix}. The length $M$ of these
lower-length Baxter operators is dictated by the powers of $x(u)$ that appear
in the perturbative expansion of the shift operator. As explicitly
seem in the three terms of~\eqref{shiftOp}: At LO, no lower-length Baxter
operators must be considered, at NLO, the operator
$B_1$ can appear, and N$^{2}$LO the operator $B_0$ can
appear.

\item[Rule 3.]

There is no explicit coupling dependence appearing
in the matrix elements. This is the simplest setup where orthogonality
can arise. To see this, one can consider the case~\eqref{linearSystem} with vanishing
residues. There are three distinct ways of
writing this linear system as a matrix equation. The first way
is~\eqref{possibility}, while the other two ways are given by
\begin{equation}
  \begin{pmatrix}
    \langle Q_S Q_J \rangle_{\mu_{1}} & \textcolor{c2}{g^2}\langle B_0[Q_S,Q_J] \rangle_{\mu_{1}} \\
    \langle Q_S Q_J \rangle_{\mu_{2}} & \textcolor{c2}{g^2}\langle B_0[Q_S,Q_J] \rangle_{\mu_{2}}
  \end{pmatrix}
  \cdot
  \begin{pmatrix}
    \Delta I_0 \\
    \frac{\textcolor{c2}{g^2}\Delta q_2^{+}}{4}
  \end{pmatrix}
  \,,
\end{equation}
or
\begin{equation}
  \begin{pmatrix}
    \langle Q_S Q_J \rangle_{\mu_{1}} & \textcolor{c2}{g^4}\langle B_0[Q_S,Q_J] \rangle_{\mu_{1}} \\
    \langle Q_S Q_J \rangle_{\mu_{2}} & \textcolor{c2}{g^4}\langle B_0[Q_S,Q_J] \rangle_{\mu_{2}}
  \end{pmatrix}
  \cdot
  \begin{pmatrix}
    \Delta I_0 \\
    \frac{\Delta q_2^{+}}{4}
  \end{pmatrix}
  \,.
\end{equation}
Due to the explicit coupling dependence of the matrix elements, the
determinants of the two matrices above are always zero at
leading order, both for equal and for distinct states, which violates
leading-order orthogonality for the Q-functions.

\end{description}

Following these rules, we construct the following matrices:
\begin{align}
  \mathcal{M}_{2,0}[Q_S,Q_J]&=
  \begin{pmatrix}
    \langle Q_S Q_J \rangle_{\mu_{1}}
  \end{pmatrix}
  \,,\nn\\
  \mathcal{M}_{2,1}[Q_S,Q_J]&=
  \begin{pmatrix}
    \langle Q_S Q_J \rangle_{\mu_{1}} & \langle (B_1 \cdot Q_S) Q_J \rangle_{\mu_{1}} \\
    \langle Q_S Q_J \rangle_{\mu_{2}} & \langle (B_1 \cdot Q_S) Q_J\rangle_{\mu_{2}}
  \end{pmatrix}
  \,,\nn\\
  \mathcal{M}_{2,2}[Q_S,Q_J]&=
  \begin{pmatrix}
    \langle Q_S Q_J \rangle_{\mu_{1}} & \langle (B_1 \cdot Q_S) Q_J \rangle_{\mu_{1}} & \langle (B_0 \cdot Q_S) Q_J \rangle_{\mu_{1}} \\
    \langle Q_S Q_J \rangle_{\mu_{2}} & \langle (B_1 \cdot Q_S) Q_J\rangle_{\mu_{2}} & \langle (B_0 \cdot Q_S) Q_J \rangle_{\mu_{2}} \\
    \langle Q_S Q_J \rangle_{\mu_{3}} & \langle (B_1 \cdot Q_S) Q_J\rangle_{\mu_{3}} & \langle (B_0 \cdot Q_S) Q_J \rangle_{\mu_{3}}
  \end{pmatrix}
  \,.
  \label{enlargedT2Lead}
\end{align}
where we allow each of the Q-functions above to have its own value
of dressing parameter~$\alpha$. Our goal is to find out whether there is a combination of
dressing parameters $\alpha$ and measures $\mu_i(u)$ that make these
enlarged matrices orthogonal.

The first matrix $\mathcal{M}_{2,0}$ has the same form as the one
previously considered~\eqref{orthoLeading}, while the other two
\emph{enlarged matrices} are new candidates for orthogonality. In
general, we will use $\mathcal{M}_{L,\ell}$ to denote the matrix of
twist-$L$ operators enlarged $\ell$ times. The orthogonality relation
will be computed by the symmetrized determinant
\begin{equation}
  \mathcal{D}_{L,\ell}[Q_S,Q_J] = \sqrt{\det\mathcal{M}_{L,\ell}[Q_S,Q_J]\,\det\mathcal{M}_{L,\ell}[Q_J,Q_S]}\,,
  \label{enlargedTreeOrtho}
\end{equation}
where taking the square root of the product is a simple way of recovering the symmetry of
exchanging the states $S \leftrightarrow J$, which was broken by the enlarged
matrices~\eqref{enlargedT2Lead} that apply the lower-length Baxter operator
to just one of the two states.

\paragraph{Leading Order.}

The fact that the coupling does not appear explicitly in the matrix
elements~\eqref{enlargedT2Lead} means that for $\ell>0$, even leading-order
orthogonality depends non-trivially on the lower-length Baxter operators.
Thus, beyond the previously discussed case of $\mathcal{M}_{2,0}$,
even at leading order, we have no guarantee that measures that make
these enlarged matrices $\mathcal{M}_{2,1}$ and $\mathcal{M}_{2,2}$
orthogonal exist. However, it turns out that such measures do exist.
They are exactly the ones considered in the twist-three
case~\eqref{twist3mu} and are given by
\begin{equation}
  \mu_1^{(0)}(u) = \frac{\pi}{2}\frac{1}{\cosh^2(\pi u)}\,,\quad \mu_2^{(0)}(u) = \pi^2 \frac{\tanh(\pi u)}{\cosh^2(\pi u)}\,,\quad \text{and} \quad \mu_3^{(0)}(u) = \frac{3\pi^3}{2}\frac{\tanh^2(\pi u)}{\cosh^2(\pi u)}\,.
  \label{mu123Tree}
\end{equation}
This results in three distinct leading-order orthogonality relations
for twist-two operators. The first one, $\mathcal{D}_{2,0}$, is exactly
the one considered in~\eqref{orthoLeading}, while the other
($\mathcal{D}_{2,1}$ and $\mathcal{D}_{2,2}$) are two new relations
that from now on we will refer to as \emph{enlarged orthogonality}.
For example, in the case of the $\ell=2$ enlarged matrices we have
\begin{gather*}
    \mathcal{M}_{2,2}[Q_4,Q_4] =
    \begin{pmatrix}
        \frac{1}{9} & 0 & 0 \\
        0 & \frac{25}{3} & 0 \\
        3 & 0 & \frac{65}{4}
    \end{pmatrix},
    \\
    \mathcal{M}_{2,2}[Q_4,Q_6] =
    \begin{pmatrix}
        0 & 0 & 0 \\
        0 & \frac{37}{25} & 0 \\
        \frac{1}{5} & 0 & \frac{343}{36}
    \end{pmatrix},
    \qquad
    \mathcal{M}_{2,2}[Q_6,Q_4] = \begin{pmatrix}
        0 & 0 & \frac{22}{15} \\
        0 & \frac{7}{15} & 0 \\
        \frac{1}{5} & 0 & 17.781
    \end{pmatrix},
\end{gather*}
whose determinants are
\begin{equation*}
   \det(\mathcal{M}_{2,2}[Q_4,Q_4]) = \frac{1625}{108}\,, \quad \det(\mathcal{M}_{2,2}[Q_4,Q_6]) =0\,, \quad \det(\mathcal{M}_{2,2}[Q_6,Q_4]) = -\frac{154}{1125}\,.
\end{equation*}

While the symmetrized determinant we defined~\eqref{enlargedTreeOrtho}
has the desired orthogonality property: $\mathcal{D}_{2,2}[Q_S,Q_J]
\propto \delta_{S,J}$, we see that the matrices themselves depend on
the ordering of the states. More precisely, we find that
$\det(\mathcal{M}_{2,2}[Q_S,Q_J]) = 0$ if $S<J$. This ordering of the
states emerges from
integrals involving lower-length Baxter operators. For example, in the case
of the two-times enlarged matrix discussed above, one of these integrals evaluates to
\begin{equation}
 \langle (B_0 \cdot Q_S) Q_J \rangle_{\mu_1} =
  \begin{cases}
  0 & \text{for}\; S>J\,, \\
  {1}/{(2S+1)} & \text{for}\; S=J\,, \\
  \frac{1}{2}(-1)^{(S-J)/2}\left(q_1^{+}(J)-q_1^{+}(S)\right) & \text{for}\; S<J\,, \\
  \end{cases}
  \label{shiftedInt}
\end{equation}
where we can clearly see the ordering in the states appearing. This is
another reason why we will always use the symmetrized determinant
$\mathcal{D}_{L,\ell}$ defined~\eqref{enlargedTreeOrtho} to compute
orthogonality among the Q-functions.

Moreover, at this order, it is easy to compare these SoV enlarged determinants
with more usual quantities, such as the Gaudin norm. We find that these
quantities are related by the following normalization factors
\begin{align}
  \mathcal{D}_{2,0}[Q_S,Q_J] & = \delta_{S,J}\frac{(2S+1)}{(2S)!} \left(Q_S\left(\sfrac{i}{2}\right)Q_S\left(-\sfrac{i}{2}\right)\right)^{-2} \times \GaudinB_2(S)
  \,,\nn\\
  \mathcal{D}_{2,1}[Q_S,Q_J] & = \delta_{S,J}\frac{(2S+1)}{(2S)!} \left(Q_S\left(\sfrac{i}{2}\right)Q_S\left(-\sfrac{i}{2}\right)\right)^{-3} \times \frac{\GaudinB_2(S)}{q_1^+(S)}
  \,,\nn\\
  \mathcal{D}_{2,2}[Q_S,Q_J] & = \delta_{S,J}\frac{(2S+1)}{(2S)!} \left(Q_S\left(\sfrac{i}{2}\right)Q_S\left(-\sfrac{i}{2}\right)\right)^{-4} \times \frac{\GaudinB_2(S)}{q_1^+(S)(q_3^+(S)-16H_3(S))}
  \,,
  \label{SoVtoBl}
\end{align}
where $H_3(S)$ is the harmonic number and $\GaudinB_{2}(S)$
is the twist-two Gaudin norm defined in~\appref{appGaudin}. There, we also
present the generalization of these relations between SoV and Gaudin
norm for operators with arbitrary twists. Note that the
relations~\eqref{SoVtoBl} are not normalization-invariant.
If one rescales the Q-function~\eqref{QFunction} by an
overall constant $\mathcal{N}$, the relation with the Gaudin norm at
leading order changes by a factor $\mathcal{N}^{L+\ell}$. One can use
this dependence to make the expressions above nicer, but it does not change the
main feature we want to emphasize: Each enlarged matrix is related to the Gaudin norm by
a different normalization factor.

\paragraph{Higher Orders.}

With the leading order established, we can now make precise how these
enlarged matrices are used to define N$^2$LO
orthogonality at twist $L=2$. The strategy remains the same: Find
perturbative corrections to the measures~\eqref{mu123Tree} that result
in orthogonality beyond leading order. To explore the vast space of
possible measures, we assume that the corrections in perturbation
theory will have the same form as in~\eqref{mutwist3loop} and~\cite{Bercini:2022jxo}, which
allows us to parametrize them in terms of a handful of constants:
\begin{equation}
  \mu_\ell^{(2)}(u) = \mu_\ell^{(0)}(u)\left(1+g^2\pi^2a_\ell \tanh^2(\pi u)+g^4\pi^4(b_\ell \tanh^2(\pi u)+c_\ell \tanh^4(\pi u)) \right)\,.
  \label{measureAnsatz}
\end{equation}
In the end, all the putative enlarged matrices~\eqref{enlargedT2Lead}
combined have 37 unknown coefficients: Nine
$\{a_1,\dots,c_3\}$ that control the perturbative corrections~\eqref{measureAnsatz} of the
three measures~\eqref{mu123Tree}, and 28 dressing parameters $\alpha$
for the Q-functions appearing in the matrix elements
of~\eqref{enlargedT2Lead}. By requiring orthogonality, \ie demanding
that the determinant of two distinct states vanishes, we were able to
uniquely\footnote{The measures are fixed up to constant terms, which can
be neglected, since at each order in perturbation theory they will
multiply factors that are vanishing due to the orthogonality relations
of the previous perturbative orders.} fix all these
coefficients.\footnote{In practice, the coefficients are fixed by
demanding orthogonality among all the states up to spin $S=10$. Then
we verified that these coefficients still imply orthogonality between
states up to spin $S=30$.}

The result can be compactly written as a set of three enlarged
matrices:
\begin{align}
  \mathcal{M}_{2,0}[Q_S,Q_J]&=
  \begin{pmatrix}
    \langle \mathcal{P}_S \mathcal{P}_J \rangle_{\mu_{1}}
  \end{pmatrix}
  \,,\nn\\
  \mathcal{M}_{2,1}[Q_S,Q_J]&=
  \begin{pmatrix}
    \langle \mathcal{P}_S \mathcal{Q}_J \rangle_{\mu_{1}} & \langle (B_1 \cdot \mathcal{P}_S) \mathcal{Q}_J \rangle_{\mu_{1}} \\
    \langle \mathcal{P}_S \mathcal{Q}_J \rangle_{\mu_{2}} & \langle (B_1 \cdot \mathcal{P}_S) \mathcal{Q}_J \rangle_{\mu_{2}}
  \end{pmatrix}
  \,,\nn\\
  \mathcal{M}_{2,2}[Q_S,Q_J]&=
  \begin{pmatrix}
    \langle \mathcal{P}_S \mathcal{Q}_J \rangle_{\mu_{1}} & \langle (B_1 \cdot \mathcal{P}_S) \mathcal{Q}_J \rangle_{\mu_{1}} & \langle (B_0 \cdot \mathcal{P}_S) \mathcal{Q}_J \rangle_{\mu_{1}} \\
    \langle \mathcal{P}_S \mathcal{Q}_J \rangle_{\mu_{2}} & \langle (B_1 \cdot \mathcal{P}_S) \mathcal{Q}_J \rangle_{\mu_{2}} & \langle (B_0 \cdot \mathcal{P}_S) \mathcal{Q}_J \rangle_{\mu_{2}} \\
    \langle \mathcal{P}_S \mathcal{Q}_J \rangle_{\mu_{3}} & \langle (B_1 \cdot \mathcal{P}_S) \mathcal{Q}_J \rangle_{\mu_{3}} & \langle (B_0 \cdot \mathcal{P}_S) \mathcal{Q}_J \rangle_{\mu_{3}}
  \end{pmatrix}
  \,.
  \label{enlargedT2}
\end{align}
Recall that $\mathcal{Q}_S(u)$ denotes Q-functions with
dressing parameter $\alpha=1$ (these are exactly the symmetrized momentum-carrying
Q-function of the QSC), while $\mathcal{P}_S(u)$ have $\alpha=0$,
\ie they are the polynomial parts of the corresponding $\mathcal{Q}_S(u)$. Meanwhile, the corrected
measures in perturbation theory are given by
\begin{align}
  \mu_1^{(2)}(u) &= \frac{\pi}{2}\frac{1}{\cosh^2(\pi u)}\left(1 + 2\pi^2g^2\tanh^2(\pi u)-\frac{g^4\pi^4}{3}\left(11\tanh^2(\pi u)-15\tanh^4(\pi u)\right)\right)\,,\nn\\
  \mu_2^{(2)}(u) &= \pi^2 \frac{\tanh(\pi u)}{\cosh^2(\pi u)}\left(1 + 3\pi^2g^2\tanh^2(\pi u)-\frac{g^4\pi^4}{3}\left(18\tanh^2(\pi u)-27\tanh^4(\pi u)\right)\right)\,,\nn\\
  \mu_3^{(2)}(u) &= \frac{3\pi^3}{2}\frac{\tanh^2(\pi u)}{\cosh^2(\pi u)}\left(1 + 4\pi^2g^2\tanh^2(\pi u)-\frac{g^4\pi^4}{3}\left(25\tanh^2(\pi u)-42\tanh^4(\pi u)\right)\right)\label{muj2Loop}\,.
\end{align}
With the dressing and measures fixed, we can now make the precise
statement of enlarged orthogonality for twist-two operators: The matrix
enlarged $\ell$ times defines orthogonality up to order
N$^\ell$LO, or equivalently
\begin{align}
  \mathcal{D}_{2,0}[Q_S,Q_J] &\propto \delta_{S,J} + O(g^2)\,,\nn\\
  \mathcal{D}_{2,1}[Q_S,Q_J] &\propto \delta_{S,J} + O(g^4)\,,\nn\\
  \mathcal{D}_{2,2}[Q_S,Q_J] &\propto \delta_{S,J} + O(g^6)\,.
\end{align}
The two-times-enlarged determinant $\mathcal{D}_{2,2}$ is our new
SoV-inspired proposal of N$^2$LO orthogonality for
twist-two operators. For example, when the two operators have spins four
and six, this two-times-enlarged matrix can be easily computed
numerically to be%
\footnote{One might not be impressed by the N$^2$LO
orthogonality exemplified, since some matrix elements are already of
order $\mathcal{O}(g^4)$, but this is an artifact of twist-two operators
and their symmetric Bethe roots. When we promote orthogonality to
operators with arbitrary twist, all matrix elements will generically start at
order~$\mathcal{O}(g^0)$.}
\begin{align*}
    &\mspace{-5mu}
    \det\mathcal{M}_{2,2}[Q_4,Q_6] =
    \\\nn &
    \begin{vmatrix}
         3.2\mathinner{\textcolor{c2}{g^4}} & -12.1 i \mathinner{\textcolor{c2}{g^2}} + 220.3 i \mathinner{\textcolor{c2}{g^4}} & 152.5\mathinner{\textcolor{c2}{g^4}}  \\
        -1.3 i \mathinner{\textcolor{c2}{g^2}} - 4 i \mathinner{\textcolor{c2}{g^4}} & -2.5 + 10.4\mathinner{\textcolor{c2}{g^2}} + 820.7 \mathinner{\textcolor{c2}{g^4}} & -62.2 i \mathinner{\textcolor{c2}{g^2}} + 873.4 \mathinner{\textcolor{c2}{g^4}}\\
        -0.2 - 3.4 \mathinner{\textcolor{c2}{g^2}} + 113.9 \mathinner{\textcolor{c2}{g^4}} & -304.3 i \mathinner{\textcolor{c2}{g^2}} - 830.7 i \mathinner{\textcolor{c2}{g^4}} & -9.5 + 0.2\mathinner{\textcolor{c2}{g^2}} + 4020.3 \mathinner{\textcolor{c2}{g^4}}
    \end{vmatrix}
    = \mathcal{O}(g^6)\,.
\end{align*}

The fact that the matrices responsible for orthogonality grow with the
order in perturbation theory is consistent with the SoV expectations.
However, contrary to expectations, the enlargement we are obtaining
does not seem to come from a weak-coupling reduction of some larger
matrix (at least we were not able to write it like that). If a simple
perturbative reduction existed, the normalization factor between the
Gaudin norm and each one of the enlarged determinants
($\mathcal{D}_{2,0}$, $\mathcal{D}_{2,1}$ and $\mathcal{D}_{2,2}$)
should be the same at leading order, and only start to differ as we go
to higher orders in perturbation theory. Instead, we observe that
the enlarged matrices have increasingly more complicated normalization
factors~\eqref{SoVtoBl} with the Gaudin norm already at
leading order.

Due to this increase in complexity, we were not
able to write the precise relation between our SoV expressions and the
Gaudin norm beyond leading order, or equivalently, the precise normalization between SoV
determinants and two-point functions. The best we were able to do, for
twist-two as well as for higher-twist operators, is to find the cases where
this two-point function vanishes. More precisely, we were able to find
matrices and measures that define orthogonality relations for the
Q-functions at each order in perturbation theory up to wrapping
corrections. This is what we turn to next.

\section{Orthogonality Before Wrapping}
\label{secOrthoAll}

\subsection{Proposal}

Before considering the weak-coupling expansion for operators of
general twist, let us first recall the Baxter equation for operators of
twist $L$ at leading order:
\begin{equation}
    \left(u+\frac{i}{2}\right)^L Q_S(u+i)+\left(u-\frac{i}{2}\right)^L Q_S(u-i) - \left(2u^L+\sum_{k=0}^{L-2}I_k(S)u^k\right)Q_S(u)=0\,.
    \label{BaxterL}
\end{equation}
Due to the appearance of $(L-1)$ integrals of motion
$I_k(S)$, the orthogonality for higher-twist operators is given by a
matrix already at leading order. Orthogonality of the Q-functions is the statement
that for any two distinct twist-$L$ states, the determinant of the
following $(L-1)\times(L-1)$ matrix vanishes:
\begin{equation}
\mathcal{M}_{L,0}[Q_S,Q_J] =
\brk*{
\begin{array}{@{}llcl@{}}
\langle \mathcal{P}_S \mathcal{P}_J \rangle_{1}      & \langle u^{1} \mathcal{P}_S \mathcal{P}_J \rangle_{1}      & \dots   & \langle u^{L-2} \mathcal{P}_S \mathcal{P}_J \rangle_{1} \\
\langle \mathcal{P}_S \mathcal{P}_J \rangle_{2}      & \langle u^{1} \mathcal{P}_S \mathcal{P}_J \rangle_{2}      & \dots   & \langle u^{L-2} \mathcal{P}_S \mathcal{P}_J \rangle_{2} \\
\hspace{1.5em} \vdots                                & \hspace{2em} \vdots                                        & \vdots  & \hspace{2.5em} \vdots \\[0.5ex]
\langle \mathcal{P}_S \mathcal{P}_J \rangle_{{L-1}}  & \langle u^{1} \mathcal{P}_S \mathcal{P}_J \rangle_{{L-1}}  & \dots   & \langle u^{L-2} \mathcal{P}_S \mathcal{P}_J \rangle_{{L-1}} \\
\end{array}
} \,,
\label{usualOrtho}
\end{equation}
where we emphasize the leading-order aspect of this result by writing
only the polynomial part $\mathcal{P}_S$ of the
Q-functions~\eqref{polyQ}, and recall that each of the brackets
above stands for the integral on the real line
\begin{equation}
  \langle f \rangle_{\ell} = \int_{-\infty}^\infty f(u) \mu_\ell(u)\,du\,,
\end{equation}
where the measures at leading order are given by
\begin{equation}
  \mu_\ell^{(0)}(u) = \frac{\ell\, \pi^\ell /2}{\cosh^2(\pi u)}\tanh^{\ell-1}(\pi u)\,.
  \label{eq:LOmeasures}
\end{equation}
To promote this result beyond leading order, for each matrix
$\mathcal{M}_{L,0}$ and integer $0\leq \ell \leq L$, we introduce the
$\ell$-times enlarged matrix
\begin{tcolorbox}[boxsep=0pt,left=0pt,right=7pt,top=-1ex,bottom=2ex]
\begin{align}
  \label{enlargedAll}
  &
  \mathcal{M}_{L,\ell}[Q_S,Q_J]=
  \\\nn & \mspace{-5mu}
  \left(
  \begin{array}{@{}l@{\;\;}c@{\;\;}l@{\;\;}l@{\;\;}c@{\;\;}l@{}}
  \textcolor{c1}{\langle \mathcal{P}_S \mathcal{Q}_J \rangle_1}          & \textcolor{c1}{\cdots} & \textcolor{c1}{\langle u^{L-2} \mathcal{P}_S \mathcal{Q}_J  \rangle_1}          & \textcolor{c2}{\langle (B_{L-1}\mathcal{P}_S) \mathcal{Q}_J \rangle_1}          & \textcolor{c2}{\cdots} & \textcolor{c2}{\langle (B_{L-\ell} \mathcal{P}_S) \mathcal{Q}_J \rangle_1} \\
  \textcolor{c1}{\langle \mathcal{P}_S \mathcal{Q}_J \rangle_2}          & \textcolor{c1}{\cdots} & \textcolor{c1}{\langle u^{L-2} \mathcal{P}_S \mathcal{Q}_J  \rangle_2}          & \textcolor{c2}{\langle (B_{L-1}\mathcal{P}_S) \mathcal{Q}_J \rangle_2}          & \textcolor{c2}{\cdots} & \textcolor{c2}{\langle (B_{L-\ell} \mathcal{P}_S) \mathcal{Q}_J \rangle_2} \\
  \hspace{1.5em}\textcolor{c1}{\vdots}                                   & \textcolor{c1}{\vdots} & \hspace{2.4em}\textcolor{c1}{\vdots}                                            & \hspace{3em}\textcolor{c2}{\vdots}                                              & \textcolor{c2}{\vdots} & \hspace{3em}\textcolor{c2}{\vdots} \\[.5ex]
  \textcolor{c1}{\langle \mathcal{P}_S \mathcal{Q}_J \rangle_{L-1}}      & \textcolor{c1}{\cdots} & \textcolor{c1}{\langle u^{L-2} \mathcal{P}_S \mathcal{Q}_J  \rangle_{L-1}}      & \textcolor{c2}{\langle (B_{L-1}\mathcal{P}_S) \mathcal{Q}_J \rangle_{L-1}}      & \textcolor{c2}{\cdots} & \textcolor{c2}{\langle (B_{L-\ell} \mathcal{P}_S) \mathcal{Q}_J \rangle_{L-1}} \\
  \textcolor{c2}{\langle \mathcal{P}_S \mathcal{Q}_J \rangle_{L}}        & \textcolor{c2}{\cdots} & \textcolor{c2}{\langle u^{L-2} \mathcal{P}_S \mathcal{Q}_J  \rangle_{L}}        & \textcolor{c2}{\langle (B_{L-1}\mathcal{P}_S) \mathcal{Q}_J \rangle_{L}}        & \textcolor{c2}{\cdots} & \textcolor{c2}{\langle (B_{L-\ell} \mathcal{P}_S) \mathcal{Q}_J \rangle_{L}} \\
  \hspace{1.5em}\textcolor{c2}{\vdots}                                   & \textcolor{c2}{\vdots} & \hspace{2.4em}\textcolor{c2}{\vdots}                                            & \hspace{3em}\textcolor{c2}{\vdots}                                              & \textcolor{c2}{\vdots} & \hspace{3em}\textcolor{c2}{\vdots} \\[.5ex]
  \textcolor{c2}{\langle \mathcal{P}_S \mathcal{Q}_J \rangle_{L-1+\ell}} & \textcolor{c2}{\cdots} & \textcolor{c2}{\langle u^{L-2} \mathcal{P}_S \mathcal{Q}_J  \rangle_{L-1+\ell}} & \textcolor{c2}{\langle (B_{L-1}\mathcal{P}_S) \mathcal{Q}_J \rangle_{L-1+\ell}} & \textcolor{c2}{\cdots} & \textcolor{c2}{\langle (B_{L-\ell} \mathcal{P}_S) \mathcal{Q}_J \rangle_{L-1+\ell}}
  \end{array}
  \right),
\end{align}
\end{tcolorbox}
where we remind that $\mathcal{P}_S$ and $\mathcal{Q}_S$ are given
by~\eqref{QFunction} with dressing parameters $\alpha=0$ and
$\alpha=1$, respectively, while $B_M$ stands for the lower-length Baxter
operator
\begin{equation}
B_M \mathcal{P}_S = \left(u+\frac{i}{2}\right)^M \mathcal{P}_S(u+i) + \left(u-\frac{i}{2}\right)^M \mathcal{P}_S(u-i) - 2u^M \mathcal{P}_S(u)\,.
\label{BLOperator}
\end{equation}
The upper left $(L-1)\times(L-1)$ part (black) is the usual
orthogonality matrix, while the new rows and columns (red)
are the enlargement to a $(L-1+\ell)\times(L-1+\ell)$ matrix via
the addition of lower-length Baxter operators and further measures.

Our main result can be stated in the following way:
\begin{tcolorbox}
The enlarged matrices $\mathcal{M}_{L,\ell}$ define twist-$L$ orthogonality at N$^\ell$LO :
\begin{equation}
  \mathcal{D}_{L,\ell}[Q_S,Q_J] \equiv
  \sqrt{\det\mathcal{M}_{L,\ell}[Q_S,Q_J]\det\mathcal{M}_{L,\ell}[Q_J,Q_S]}
  \propto \delta_{S,J}+\mathcal{O}(g^{2(\ell+1)})
  \label{claim}
\end{equation}
Where the measures responsible for this orthogonality are given by
\begin{align}
  \mu_\ell(u) & = \frac{\ell \pi^\ell /2}{\cosh^2(\pi u)}\tanh^{\ell-1}(\pi u)\,e^{\nu_\ell(u)}\,,
  \label{measuresAll} \\
  \nu_\ell(u) & = \sum_{n=1}^\infty\Bigg(\!
    \binom{2n-1}{n}\frac{\ell+1}{n}
    - (g\pi)^2\binom{2n}{n}\frac{(\ell+1)(1+{4n}/{3})-1}{n+1}
  \Bigg) (g\pi)^{2n} \tanh^{2n}(\pi u)\,.
  \nonumber
\end{align}
\end{tcolorbox}
We emphasize that the measures $\mu_l$ above are twist-independent,
depending only on the row of the matrix in which they appear. Note also
that just as in the twist-two case, the enlarged matrices will be
sensitive to the ordering of the states:
$\det(\mathcal{M}_{L,\ell}[Q_S,Q_J]) = 0$ only if $S<J$, thus it is
necessary to consider the symmetrized determinant
$\mathcal{D}_{L,\ell}$ to define the orthogonality~\eqref{claim} among
the Q-functions. Note also that for any ordering of the states, when
one evaluates the determinant of the enlarged
matrices~\eqref{enlargedAll}, it is possible to factor out the measures and
Q-functions to write the orthogonality relation in a very suggestive
form:
\begin{equation}
    \det(\mathcal{M}_{L,\ell}) = \int \left(\prod_{n=1}^{L+\ell-1} du_n \,  \mu_n(u_n)\right) \times \left(\prod_{n=1}^{L+\ell-1}\mathcal{P}_S(u_n)\right) \times \overleftarrow{\mathbb{V}} \times  \left(\prod_{n=1}^{L+\ell-1}\mathcal{Q}_J(u_n)\right)
    \,,
    \label{eq:detStateLinear}
\end{equation}
where $\overleftarrow{\mathbb{V}}$ is an operator that acts on the
Q-functions $\mathcal{P}_S$ on the left defined as
\begin{equation}
   \overleftarrow{\mathbb{V}} =
   \left|
   \begin{array}{ccllcl}
     1 & \dots & u_1^{L-2} & \overleftarrow{B}_{L-1}(u_1) & \dots & \overleftarrow{B}_{L-\ell}(u_1) \\
     1 & \dots & u_2^{L-2} & \overleftarrow{B}_{L-1}(u_2) & \dots & \overleftarrow{B}_{L-\ell}(u_2) \\
     \vdots & \vdots & \hspace{.7em} \vdots & \hspace{2em} \vdots & \vdots & \hspace{2em} \vdots \\[0.5ex]
     1 & \dots & u_{L+\ell-1}^{L-2} & \overleftarrow{B}_{L-1}(u_{L+\ell-1}) & \dots & \overleftarrow{B}_{L-\ell}(u_{L+\ell-1})
   \end{array}
   \right|\,.
\end{equation}
When there is no enlarging, $\overleftarrow{\mathbb{V}}$ is no longer
an operator acting on the Q-functions, but rather the familiar
Vandermonde determinant that appears in $\alg{sl}(2)$ SoV
orthogonality at leading order.
The expression~\eqref{eq:detStateLinear} highlights that
$\det(\mathcal{M}_{L,\ell})$ is a bilinear form in the wavefunctions (products of
Q-functions), which is a necessary requirement for any scalar product
on the space of states.

Despite the non-democratic nature of the two states in this
formula, such expressions are familiar from the separation of
variables of higher rank $\alg{sl}(n)$ spin chains
\cite{Cavaglia:2019pow}, whose scalar products also feature
Q-functions entering in a non-democratic fashion. The Q-functions for
one state enter as simple products, while the others are acted on by
certain shift operators. In those setups, this lack of symmetry
between the two states can be traced to the fact that one state lives
in the fundamental representation, while the other is
anti-fundamental, see \cite{Gromov:2019wmz}. In our set-up, the origin
of the asymmetry is less clear.

\subsection{Properties}

\paragraph{Derivation.}

There is a lot to unpack in this result, so let us start by addressing
how the matrices~\eqref{enlargedAll} and measures~\eqref{measuresAll}
were obtained. Similarly to the twist-two case at N$^2$LO,
we started constructing the enlarged matrices following the three
rules:
\begin{description}
  \item[Rule 1.] All Q-functions are of the form~\eqref{QFunction}, with general parameter $\alpha$.
  \item[Rule 2.] Matrix elements are built only of Q-functions and lower-length Baxter operators.
  \item[Rule 3.] The matrix elements have no explicit coupling dependence.
\end{description}
For the measure, we followed~\cite{Bercini:2022jxo} and the twist-two
result~\eqref{muj2Loop}, by also assuming that the
perturbative corrections to the measures can be parametrized in the
following way
\begin{equation}
    \mu_\ell(u) = \mu_\ell^{(0)}(u)\sum_{n=0}^{\infty}\sum_{m=0}^{n}(g\pi)^{2n}a_{n,m}\tanh^{2m}(\pi u)\,.
    \label{measureAnsatzAll}
\end{equation}
By demanding orthogonality among different states at a fixed order in
perturbation theory, one can completely fix the unknown coefficients
$a_{n,m}$ of the measures~\eqref{measureAnsatzAll}.%
\footnote{Some coefficients of the measures multiply terms that are
zero via lower-order orthogonality. These coefficients, of course, are
not fixed. They are completely irrelevant for orthogonality and
only affect the relative factor between the norm of the states and the
Gaudin norm, which we do not make precise in this work.}
We then verified
that these fixed measures yield orthogonality for a
plethora of further states (where we avoid problematic cases, see below).%
\footnote{For higher twist at higher loop orders,
computing these enlarged determinants becomes numerically challenging,
but it is relatively easy to verify the orthogonality evaluations at
N$^5$LO for all states up to twist five and spin ten,
which are already 178 different states.}
In this way, the proposal~\eqref{claim} is not derived from some
underlying principles, instead at
each order in perturbation theory, we can \emph{prove it by
exhaustion}. We provide a \mathematica notebook that computes
these determinants~\eqref{claim} for any two states at any
perturbative order before wrapping corrections, there we also present
non-trivial examples, such as the N$^{5}$LO orthogonality
between two twist-five states of spins six and eight.

After considering orthogonality up to N$^5$LO in the
weak-coupling expansion, we started to notice patterns emerging in the
coefficients of the measures. Namely, we observed that the measure
exponentiates, and that the coefficients of $\tanh(\pi u)$ can be
written in the closed form presented in~\eqref{measuresAll}. To give
further evidence that this pattern keeps holding at higher orders in
perturbation theory, we then checked that these measures make twist-six
states of spins two and four orthogonal at
N$^6$LO.%
\footnote{The
numerical evaluation becomes heavy at this loop order -- orthogonality
for operators of twist-six at N$^{6}$LO is the
determinant of an $11\times11$ matrix. Hence, we restricted ourselves to
spins two and four for this check.}
This supports the idea that the patterns of the
measure persist at higher orders in perturbation theory, allowing us
to conjecture the all-loop expression~\eqref{measuresAll}.

\paragraph{Lack of Orthogonality for Equal-Spin States.}

To make our expressions precise, we must address what we mean by
\emph{avoiding problematic cases}. Note that in our main
result~\eqref{claim} we write a Kronecker delta in the \emph{spins}
rather than in the \emph{states}. For twist-two operators, the two
conditions are interchangeable, since for each value of spin, there is
just a single state. On the other hand, at higher twists, different
operators of the same spin start appearing. The major
drawback of our proposal~\eqref{claim} is that \emph{states of equal
twist and equal spin are not orthogonal!}
For example, twist-three operators have five
states with spin 12; these five states are orthogonal to all other
twist-three states, but they are not orthogonal among themselves -- this
is also exemplified in the attached \mathematica notebook.

This lack of orthogonality among states of the same spin is deeply related
with the integrals involving the lower-length Baxter operators,
which introduce orderings in the spins of the
states~\eqref{shiftedInt}. Such integrals become more and more
complicated as we increase the
twist, the length of the lower-length Baxter operator, and the perturbative order.
This means that we quickly lose the ability to write closed expressions
like~\eqref{shiftedInt}, and even in the simplest case
of twist-three operators at leading order, we lose track of the analytic
spin dependence in the enlarged matrices. Without knowing precisely
how the spins of the states affect the enlarged matrices, there is no way to
better understand or even solve this problem of a lack of
orthogonality for operators with the same values of spin.

\paragraph{Orthogonality for States of Different Twists.}

Notably, provided we avoid states with the same spin, our
proposal~\eqref{claim} also makes states of \emph{different twists}
orthogonal to each other. This is an important property of any putative finite-coupling SoV
formalism. The SoV formalism for rational spin chains has the length
$L$ fixed. But at finite coupling in $\mathcal{N}=4$ SYM, the twist is
not a quantum number, hence states with different twists will mix. The
relevant $\alg{psu}(2,2|4)$ quantum number is R-charge, which becomes
the twist $L$ in the weakly coupled $\alg{sl}(2)$ sector that we
consider here.
Hence, two states with different
R-charge must be orthogonal, and so must be orthogonal in the SoV
formalism, which is what we observe. For example, the leading-order
orthogonality of a twist-two state with spin four and a twist-three
state of spin six is given by
\begin{equation*}
\det(\mathcal{M}[Q_{L=2,S=4},Q_{L=3,S=6}]) =
    \begin{vmatrix}
 0.548 & 1.014 & -11.226 \\
 3.436 & 6.360 & -70.433 \\
 16.199 & 30.034 & -332.084
\end{vmatrix} = 0\,.
\end{equation*}

Without including wrapping effects, our proposal is far from capturing such
finite-coupling mixing results, but it does encode the perturbative
version of that (orthogonality of operators with different lengths) in
a very simple way.
To see this, note that in the lower-length Baxter
operator~\eqref{lowerBaxter}, we only included the universal term~$u^M$ of the
transfer matrix of the full Baxter operator~\eqref{BaxterL}. This is the
\emph{only term} of the transfer matrix that matters for the enlarged
orthogonality. We could have included the complete transfer
matrix of~\eqref{BaxterL}, but then all lower powers of $u$ can be removed
from the enlarged matrices~\eqref{enlargedAll} via linear
transformations of the columns.
Therefore, for the purpose of orthogonality, the lower-length Baxter~\eqref{lowerBaxter}
with only the universal transfer matrix term and the full
Baxter operator~\eqref{BaxterL} with all powers of~$u$ are equivalent.
We opt to write the lower-length Baxter operators with
only the universal term to make manifest that
\emph{all state-dependent information} of our enlarged orthogonality
enters solely through the Q-functions.

When we consider the overlap of states with different twists, the
state with the smallest twist will be annihilated by one of the
lower-length Baxter operators. This produces entire columns of zeros in the
enlarged matrices, making their determinants vanish, and the states
trivially orthogonal. This also provides a nice physical
interpretation for each of the enlargings we consider: Matrices with
no enlarging are the usual orthogonality that makes twist-$L$ states
orthogonal among themselves, the first enlarging introduces $B_{L-1}$,
making twist-$L$ states orthogonal with twist-$(L-1)$ states, the second
enlarging introduces $B_{L-2}$, making them orthogonal with
twist-$(L-2)$ states, and so on, until we reach the vacuum state with zero twist.

\paragraph{Wrapping.}

One important aspect of our proposal~\eqref{claim} is that we cannot
enlarge the matrices forever; the matrix $\mathcal{M}_{L,\ell}$ can
only be enlarged \emph{at most $L$ times}. This means that twist-two can
be enlarged two times, twist-three enlarged three times, and so on. Since
each enlargement corresponds to orthogonality up to a particular order
in perturbation theory, an equivalent statement is that twist-two
operators can be made orthogonal up to N$^2$LO, twist-three
operators up to N$^3$LO, and so on. These are exactly
the orders in perturbation theory where wrapping corrections for
these operators start to contribute. In other words, the enlarged matrices define
perturbative orthogonality up to wrapping order.

\paragraph{Result.}

We can summarize the claim~\eqref{claim} and all the
properties discussed above as the following statement:

\begin{tcolorbox}
  For any two states of twists $L$ and $M$, with
  $L\geq M$, excluding states of equal twist and equal spin, there is
  a set of universal measures~\eqref{measuresAll}
  and $L+1$ enlarged matrices~\eqref{enlargedAll}:
  $\{\mathcal{M}_{L,0},\mathcal{M}_{L,1},\dots,\mathcal{M}_{L,L}\}$,
  whose determinants~\eqref{claim} vanish at
  $\{\text{LO},\text{NLO},\dots,\text{N}^L\text{LO}\}$, respectively.
\end{tcolorbox}

\paragraph{Finite Coupling Measure.}

\begin{figure}
    \centering
    \includegraphics[width=1\linewidth]{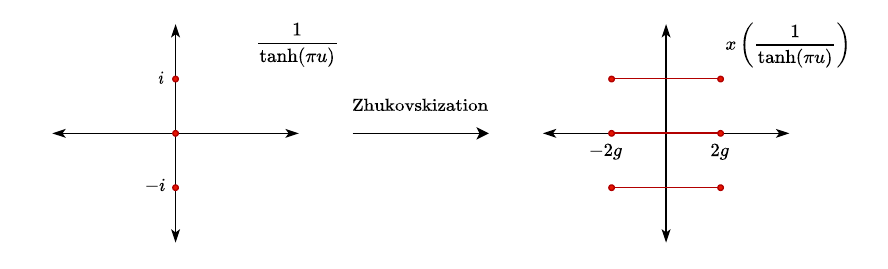}
    \caption{Zhukovskization of the measure opens up the $i$-periodic
    poles into an infinite ladder of cuts between $[-2g,2g]$.}
    \label{figZhuk}
\end{figure}

It is easy to perform the infinite sums appearing in the perturbative
measure~\eqref{measuresAll} and write them in a closed expression in
terms of the 't~Hooft coupling:
\begin{align}
  \label{finiteMu}
  \mu_\ell(u) &= \frac{\ell (2\pi)^\ell\tanh^{\ell-1}(\pi u)}{\cosh^2(\pi u)}
  \times\left(\frac{1}{1+\sqrt{1+(2\pi g\tanh(\pi u))^2}} \right)^{\ell+1}
  \times \\ \nonumber & \mspace{-30mu} \times
  \exp\Bigg[g^2\pi^2\bigg(\ell+
  \frac{2(\ell+4)}{3}\frac{1}{1+\sqrt{1+(2\pi g\tanh(\pi u))^2}}-\frac{4(\ell+1)}{3}\frac{1}{\sqrt{1+(2\pi g\tanh(\pi u))^2}}\bigg)\Bigg].
\end{align}
The first term is the usual tree-level measure, while the other two
terms come from summing the two binomials in~\eqref{measuresAll}. With
the help of Zhukovsky variables, we can rewrite the square
roots of the above expression in a relatively compact
finite-coupling expression
\begin{tcolorbox}[boxsep=0pt,left=0pt,right=5pt,top=-1ex,bottom=1.2ex]
\begin{equation}
  \mu_\ell(u) = \frac{\ell}{2\pi\sinh^2(\pi u)}\frac{1}{x(\tau)^{\ell+1}}
  \exp
  \brk[s]*{
  {g^2\pi^2\left(\ell +\frac{4+\ell}{3}\frac{\tau}{x(\tau)}+\frac{4(\ell+1)}{3}\frac{1}{1-2{x(\tau)}/{\tau}}\right)}\,
  },
  \label{measureFinite}
\end{equation}
\end{tcolorbox}
where $\tau(u) = {1}/\brk{\pi\tanh(\pi u)}$. This Zhukovsky map
of the measure is similar to the Zhukovski\-zation procedure considered
in~\cite{Cavaglia:2021mft}, its effect is to promote some of the $i$-periodic
poles of the measure to $i$-periodic cuts, as depicted
in~\figref{figZhuk}.

Spoiling the beautiful result in terms of Zhukovsky
variables~\eqref{measureFinite} is the exponential, which depends
explicitly on $\tau(u)$ and the coupling. Perhaps, by correctly
incorporating wrapping effects into the SoV framework, one could
generate terms that combine with this exponential and result in
an expression for the measure where the coupling dependence enters
only via the Zhukovsky variables. Obtaining such finite-coupling
measures is the holy grail of the SoV formalism in $\mathcal{N}=4$
super Yang--Mills, and is what is missing to elevate it to the same
standards as the Quantum Spectral Curve.

Unfortunately, our proposal~\eqref{claim} is not there yet. It does
not make states with the same spin orthogonal, and its leading-order
relation with the Gaudin norm changes with each successive
enlargement (see~\appref{appGaudin}). To resolve these problems, we
would need to have complete analytic control over these enlarged orthogonality
expressions at N$^2$LO and beyond, which we are
currently lacking. In the next section, we provide an
\emph{alternative} enlarged orthogonality formulation, where we do
have such analytic control, and we comment on how some of these
problems can be resolved.

\section{Orthogonality With Residues}
\label{WayOut}

\subsection{SoV with One Q-Function}

To develop this alternative formulation, we go back to twist-two
operators. Our starting point will be
equation~\eqref{twoLoopSelfAdj}, but we will make it more democratic in
the Q-functions by symmetrizing it in the states:
\begin{equation}
  \Delta I_0 \langle \mathbb{Q}_S \mathbb{Q}_J \rangle_\mu + g^4\frac{\Delta q_2^{+}}{4} \langle  B_0[\mathbb{Q}_S,\mathbb{Q}_J] \rangle_\mu + \texttt{res}_\mu = 0\,,
  \label{twoLoopSelfAdjAnti}
\end{equation}
with $B_M[F,G] \equiv (B_M \cdot F)G + (B_M \cdot G)F$ being the symmetric
action of lower-length Baxter operators~\eqref{lowerBaxter}, and the
Q-functions are given by~\eqref{QFunction} with $\alpha={1}/{2}$
as dressing parameters. The residues that are picked by the
contour shifts are given by
\begin{align}
  \texttt{res}_\mu &= \Res_{u={i}/{2}}\left[\mathbb{Q}_S(u)\left(\brk{x^-}^2-\frac{g^4}{4}\brk1{q_2^{+}(S)+q_2^{+}(J)}\right)\mathbb{Q}_J(u-i)\mathinner{\mu(u)}\right] \nn\\
  & \quad +\Res_{u=-{i}/{2}}\left[\mathbb{Q}_S(u)\left(\brk{x^+}^2-\frac{g^4}{4}\brk1{q_2^{+}(S)+q_2^{+}(J)}\right)\mathbb{Q}_J(u+i)\mathinner{\mu(u)}\right]\,.
  \label{resExplicitMain}
\end{align}
As previously recalled in~\secref{secTwist2}, the guiding idea of the
Separation of Variables formalism is to find as many measures with
vanishing residues as integrals of motion and
conserved charges appear in the equation.

For any given measure, it is possible to compute the
residue~\eqref{resExplicitMain} in terms of the conserved charges
$q_n^{\pm}(S)$ and $q_n^{\pm}(J)$, \emph{without} having to specify
their on-shell values in terms of Bethe roots. Since these charges are
independent, by demanding that the coefficient of each of them is
zero, we can find measures with vanishing residues in a fully
off-shell way. We explain this procedure further in~\appref{appResidues}, and
it trivializes the searching for SoV measures. For example, in the
twist-two case~\eqref{resExplicitMain}, we find that up to
N$^2$LO, there exists a single measure with vanishing
residues, given by
\begin{equation}
  \rho_1(u) = \frac{\pi/2}{\cosh^2(\pi u)}\brk*{1-g^2\pi^2(1-3\tanh^2(\pi u))+\frac{2}{3}g^4\pi^4(2-15\tanh^2(\pi u)+15\tanh^4(\pi u))}.
  \label{mu1}
\end{equation}
Since the conserved charge $q_2^{+}$ appears only at
N$^2$LO, up to $\text{NLO}$ there is a single integral
of motion $I_0$, and therefore a single measure is sufficient to define
orthogonality. In other words, the measure~\eqref{mu1} defines
next-to-leading order orthogonality%
\footnote{The first perturbative
order to~\eqref{mu1} is exactly the same as the one proposed
in~\cite{Bercini:2022jxo}. The second perturbative order of the
measure~\eqref{mu1} is equal to the first line of equation (14)
in~\cite{Bercini:2022jxo}, which is the state-independent part of the
measure considered there.}
\begin{equation}
    \mathbb{M}_{S,J}^{(1)} =
    \begin{pmatrix}
        \langle \mathbb{Q}_S \mathbb{Q}_J \rangle_{\nu_1}
    \end{pmatrix} \,, \quad
    \det \left(\mathbb{M}_{S,J}^{(1)}\right) \propto \delta_{S,J} + \mathcal{O}(g^4)
    \,.
    \label{oneLoopOrtho}
\end{equation}
Another way of writing this measure is via the
\emph{Zhukovskization} procedure~\cite{Cavaglia:2021mft}. By convoluting the leading
order term with the Zhukovsky variable, we can recover all its
perturbative expansion via the simple contour integral
\begin{equation}
  \rho_1(u) \equiv \oint \frac{dv}{2\pi i} \frac{\pi/2}{\cosh^2(\pi (u-v))} \frac{1}{x(v)}\,,
  \label{eqZhukovskyzation}
\end{equation}
which is a similar way of promoting these measures beyond
perturbation theory, by opening the poles into cuts as depicted
in~\figref{figZhuk}. Although exciting, we
will not consider finite-coupling aspects any further, since the
orthogonality we develop below, so far, is only applicable at
N$^2$LO in the context of twist-two operators.

Two measures with vanishing residues are needed to define
N$^2$LO orthogonality, but only one measure that enjoys
such a property exists in the setup~\eqref{twoLoopSelfAdjAnti}. As
explained before, there are many directions
one can pursue in the SoV framework to address this problem.
In~\secref{secOrthoAll}, we took a more numerical approach,
bypassing all residue computations, and searching directly for
orthogonal measures. Here, we follow a more analytical approach by
focusing on other ways that the residues can yield orthogonality for
the Q-functions.

Our explorations will rely on the following rather simple observation: Since the
goal is to have as many measures as integrals of motion and conserved
charges appear in the equation, the residues do not
necessarily need to be zero, but themselves can also given by a
combination of these (differences of) integrals of motion and conserved charges.
More explicitly, in the equation~\eqref{twoLoopSelfAdjAnti},
we can extend our search from measures with vanishing residues to
measures whose residues are given by the difference $\Delta q_2^{+}$ of
the conserved charge~$q_2^{+}$, multiplied by integrals with
state-independent measures.

\subsection{SoV with Two Q-Functions}
\label{sec:Q2}

Even though we enlarged our search for measures, it turns out that no
measure generates residues proportional to \emph{only} $\Delta q_2^{+}$,
other conserved charges $q_{n}^+$ also appear. But there is a way out
of this as well. In all higher-rank integrable models, more than one
type of Q-functions that solve the Baxter equation appear in the SoV orthogonality
relation~\cite{Cavaglia:2019pow,Gromov:2019wmz,Gromov:2020fwh,Gromov:2022waj}.
However, in rank-one sectors such as $\alg{sl}(2)$, the
second solution to the Baxter equation often plays no role. As we
explain below, by including this second Q-function, we will be able
to consider another measure $\rho_2(u)$, whose residues are proportional
only to the conserved charge $\Delta q_2^{+}$. This will allow us to close
the linear system~\eqref{twoLoopSelfAdjAnti} and define a new SoV-type
expression for N$^2$LO orthogonality for twist-two
operators.

One main difference between the two Q-functions lies in their different
large-$u$ asymptotic behavior: While the Q-function we have
considered so far has asymptotics $Q^{(1)}_S(u)\sim u^S$, the
second solution behaves as $Q^{(2)}_S(u)\sim u^{-S-1}$.
Both objects can be equally easily computed via the perturbative
Quantum Spectral Curve~\cite{Marboe:2014gma}. For example, the two
Q-functions for the twist-two operator of spin two at leading order are
\begin{align}
  Q^{(1)}_{S=2}(u) &= \frac{1}{4}-3u^2 \,,\label{eq:Q1}\\
  Q^{(2)}_{S=2}(u) &= -3iu + \left(\frac{1}{4}-3u^2\right)\psi_{1}\left(\frac{1}{2}+iu\right)\label{eq:Q2}\,.
\end{align}
Both Q-functions satisfy the same Baxter equation \eqref{baxterEqLead}, and so the Baxter equation for any solution $Q(u)$ can be written as the following trivial determinant
\begin{equation}
  \left|
  \begin{array}{lll}
        Q(u+i) & Q(u) & Q(u-i) \\
        Q^{(1)}(u+i) & Q^{(1)}(u) & Q^{(1)}(u-i) \\
        Q^{(2)}(u+i) & Q^{(2)}(u) & Q^{(2)}(u-i) \\
  \end{array}
  \right|=0\,.
\end{equation}
Expanding this determinant and comparing with the usual Baxter
equation \eqref{baxterEqLead} then fixes the Wronskian at
leading-order to be
\begin{equation}
    Q^{(1)}_{S}\left(u+\frac{i}{2}\right)Q^{(2)}_{S}\left(u-\frac{i}{2}\right)-Q^{(1)}_{S}\left(u-\frac{i}{2}\right)Q^{(2)}_{S}\left(u+\frac{i}{2}\right) = \frac{1}{u^2}\,.
    \label{wronsk}
\end{equation}

In this section, we will use a different overall normalization for the
Q-functions~\eqref{QFunction} than we use elsewhere in the
paper. This normalization makes the residue computations
simpler, as explained in~\appref{appResidues}, and is achieved by
\begin{equation}
    \hat Q^{(1)}_S(u) = \frac{1}{\mathcal{P}_S\left(i/2\right)}Q^{(1)}_S(u)
    \,.
\end{equation}
Note that choosing this normalization automatically adds the factor
$\mathcal{P}_S\left(i/2\right)$ to the second Q-function, since their
product must result in the Wronskian relation~\eqref{wronsk},
\begin{equation}
\hat Q^{(2)}_S(u)= \mathcal{P}_S(i/2)Q^{(2)}_S(u)
\,.
\end{equation}
This
normalization is precisely the one that is presented in~\eqref{eq:Q1}
and~\eqref{eq:Q2} for the spin-two operator.

Since $Q_S^{(2)}$ is a solution to the same Baxter equation as
$Q^{(1)}_S$, it also satisfies the same
equation as~\eqref{twoLoopSelfAdjAnti}, explicitly:
\begin{equation}
  \Delta I_0 \langle \mathbb{Q}^{(2)}_S \mathbb{Q}^{(2)}_J \rangle_\mu + g^4\frac{\Delta q_2^{+}}{4} \langle  B_0[\mathbb{Q}^{(2)}_S,\mathbb{Q}^{(2)}_J] \rangle_\mu + \texttt{res}_\mu = 0\,.
  \label{twoLoopSelfAdjAnti2}
\end{equation}
However, the different asymptotics of this second Q-function allows us to search for periodic
measures that do not decay exponentially at infinity. In fact, we
find that the measure
\begin{equation}
  \rho_2(u) = \pi \tanh(\pi u)
  \label{mu2}
\end{equation}
in the equation~\eqref{twoLoopSelfAdjAnti2} for $Q^{(2)}_S$, generates residues that
are given by
\begin{equation}
  \texttt{res}_{\rho_2} = \frac{\Delta q_2^{+}}{4}\times \textcolor{blue}{1} =  \frac{\Delta q_2^{+}}{4} \times \mathcal{P}_S\left(\frac{i}{2}\right)\mathcal{P}_J\left(\frac{i}{2}\right)
  \,,
  \label{resmu2}
\end{equation}
where we explicitly wrote the number one in terms of the Q-functions,
to make manifest how changing the overall normalization of the
Q-functions changes our expressions.

Thus, using the two Q-functions $Q_S^{(1)}$ and $Q_S^{(2)}$, and the
two measures~\eqref{mu1} and~\eqref{mu2}, we can write the following
linear system, where the \rhs emphasizes to which perturbative
order each equation holds:
\begin{equation}
  \left\{ \begin{aligned}
    \Delta I_0 \times \langle \hat{\mathbb{Q}}_S^{(1)} \hat{\mathbb{Q}}_J^{(1)} \rangle_{\rho_{1}} &+ \frac{\Delta q_2^+}{4} \times \textcolor{c2}{g^4} \langle B_0[\hat{\mathbb{Q}}^{(1)}_S,\hat{\mathbb{Q}}^{(1)}_J] \rangle_{\rho_{1}} & = \mathcal{O}(g^6)\\
    \Delta I_0 \times \langle \hat{\mathbb{Q}}_S^{(2)} \hat{\mathbb{Q}}_J^{(2)} \rangle_{\rho_{2}} &+ \frac{\Delta q_2^+}{4} \times \textcolor{blue}{1} & = \mathcal{O}(g^2)
  \end{aligned} \right.
  \,,
  \label{linearSystem2}
\end{equation}
which we can transform into the matrix equation
\begin{equation}
  \mathbb{M}^{(2)}_{S,J}
  \begin{pmatrix}
    \Delta I_0 \\ \frac{\Delta q_2^+}{4}
  \end{pmatrix}
  =
  \begin{pmatrix}
    \order{g^6} \\ \order{g^2}
  \end{pmatrix}
  \,,\qquad
  \mathbb{M}^{(2)}_{S,J} =
  \begin{pmatrix}
    \langle \hat{\mathbb{Q}}^{(1)}_S \hat{\mathbb{Q}}^{(1)}_J \rangle_{\rho_{1}} & \textcolor{c2}{g^4}\langle B_0[\hat{\mathbb{Q}}^{(1)}_S,\hat{\mathbb{Q}}^{(1)}_J] \rangle_{\rho_{1}} \\
    \langle \hat{\mathbb{Q}}^{(2)}_S \hat{\mathbb{Q}}^{(2)}_J \rangle_{\rho_{2}} & \textcolor{blue}{1}
  \end{pmatrix}\,.
  \label{matrixQ1Q2}
\end{equation}
While the equation as written looks normalization dependent, it is easy to restore the normalization invariance by
rewriting the number one in the lower-right in terms of Q-functions as
in~\eqref{resmu2} and noticing that the overall $\mathcal{P}_S(i/2)\mathcal{P}_J(i/2)$ factors out.

One might be worried that perturbative corrections for the measure
$\rho_2(u)$, residues, and Q-functions $Q^{(2)}_S$ need to be
considered in the bottom row. But, as we will explain now, such
corrections do not matter for N$^2$LO orthogonality.

If the two states are different, the first matrix element is already order
$\mathcal{O}(g^4)$, as shown in~\eqref{oneLoopOrtho}. Thus, the entire
first row of the matrix is of order $\mathcal{O}(g^4)$, which washes
away any perturbative corrections that the bottom row might have when
computing the determinant. The matrix~\eqref{matrixQ1Q2} is another
SoV-type representation for
N$^2$LO orthogonality for twist-two operators%
\footnote{To be completely sure, we verified the
orthogonality relation~\eqref{newOrtho} for states up to spin $S=60$.}
\begin{equation}
  \det\left(\mathbb{M}^{(2)}_{S,J}\right) \propto \delta_{S,J}
  +\mathcal{O}(g^6)
  \,.
  \label{newOrtho}
\end{equation}
If the states are equal, the first matrix element is order
$\mathcal{O}(g^0)$, and therefore corrections to $Q^{(2)}_S$, measures
and residues must be taken into account to make precise the connection
between the Gaudin norm and the SoV proposal~\eqref{newOrtho}.
Nonetheless, the fact that the lower-right matrix element is $1+\mathcal{O}(g^2)$
guarantees that, despite the perturbative corrections, the enlarged
matrix $\mathbb{M}_{S,J}^{(2)}$ and the original-size matrix
$\mathbb{M}_{S,J}^{(1)}$ will have the \emph{same} normalization
factor with the Gaudin norm at leading order.\footnote{Better
understanding these corrections could shine light on the mysterious
reciprocity-like normalization factor $\mathbb{J}$ between SoV and
Gaudin norm observed in~\cite{Bercini:2022jxo}.}

The explicit coupling dependence on the matrix elements makes clear
the difference between the two SoV type expressions for twist-2
proposed in this work: \eqref{enlargedT2} and \eqref{newOrtho}. Before
we had no terms with explicit coupling dependence, which led to an
obscure normalization with the Gaudin norm. Here, the residues enter
the enlarged matrix with explicit coupling dependence, resulting in a
more transparent connection with the Gaudin norm at leading order.

The drawback of this proposal is that is very difficult to search for
measures whose residues are proportional to conserved charges, especially for the
case involving $Q^{(2)}_S$, whose residues we do not know how to write
in a off-shell way like $Q^{(1)}_S$. In practice, what we do is to
compute the residues on-shell for many different states, and then try
to recast them as a state-independent combination of conserved
charges.

This proportionality is only unambiguous if we have a precise map
between the conserved charges and the Q-functions. Otherwise, any
state-dependent change in the normalization of the Q-functions would
change the proportionality with the conserved charges. Such maps exist
for $Q^{(1)}_S$, for any coupling and any twist,
see~\eqref{eq:qContourIntegrals}. But for $Q^{(2)}_S$ we were only
able to find the
relation of this Q-function with $q_2^{+}$ for the case of twist-2
operators at leading order. Without such maps for higher twists and
higher orders in perturbation theory, the concept of proportionality
becomes ill-defined, and we were not able to push these ideas for more
complicated cases than twist-2 operators at N$^2$LO. The
proposal~\eqref{newOrtho} must be understood as a \emph{proof of
concept} to call attention to the fact that SoV explorations can
include more than one type of Q-functions (even in rank-1 sectors) and
show that it is possible to include non-vanishing residues in the SoV
framework.

\section{Discussion}
\label{Discussion}

In this work, we proposed a Separation of Variables type expression for
orthogonality in the $\alg{sl}(2)$ sector of $\mathcal{N}=4$ super
Yang--Mills theory, valid
to all orders in the weak-coupling expansion prior to wrapping
corrections. More precisely, we defined a set of ever-growing
matrices~\eqref{enlargedAll}, together with a set of finite-coupling
measures~\eqref{finiteMu}, that make two states of non-equal spins
orthogonal up to the perturbative order at which wrapping corrections
start contributing.

Besides not capturing orthogonality for states of equal spin, each
successive enlarging changes the proportionality factor with the Gaudin
norm, which makes it difficult to precisely establish the connection
between the SoV determinants proposed here and the two-point functions
of operators.

We attempted to resolve these issues by generalizing the properties
of the residues and considering both type of Q-functions for the $\alg{sl}(2)$
sector, the latter of which are already a necessary ingredient in the
SoV expressions of other
models~\cite{Ambrosino:2023dik,Ambrosino:2024prz,Fateev:2009jf,Vegh:2023snc,Litvinov:2024riz,Artemev:2025cev}.
The outcome was another
SoV-type expression~\eqref{newOrtho},
which does not change the normalization factors with the Gaudin norm.
So far, we could not generalize that orthogonality expression to operators of arbitrary
twist at higher orders in perturbation theory, which would be required
to check whether the orthogonality among operators of equal spin is resolved as well.

We believe that the general lessons drawn from the two SoV-inspired
approaches we developed in this work could serve as guidelines for
future explorations involving the Separation of Variables formalism in
$\mathcal{N}=4$ super Yang--Mills theory as well as other integrable models.
Let us review the salient points:

\paragraph{1. Dressings.}

In perturbation theory, the transfer matrix
has non-polynomial parts, whose coefficients are controlled by the
dressing $e^{\alpha\cdot\sigma(u)}$ of the Q-functions.
There is no apparent conceptual problem
in considering the resulting extra poles in the SoV formalism, but to our
knowledge, thus far no results have accounted for such poles. Our
proposal offers a simple way of removing these terms: \emph{Make the
non-polynomial part of the transfer matrix universal}. Which is
precisely what the QSC-inspired dressing~\eqref{dressing} does in the
asymptotic Baxter equation. Identifying whether these
dressed Q-functions correspond to natural objects in the context of
full QSC -- which is important in order to understand how to
incorporate wrapping effects -- remains an outstanding question.

\paragraph{2. Q-functions.}

The second orthogonality formulation we
proposed implies that even in rank-one sectors such as $\alg{sl}(2)$, all
different types of Q-functions that solve the Baxter equation should be
considered, as is the case for higher-rank integrable models \cite{Ryan:2018fyo,Cavaglia:2019pow}.
Moreover, in both orthogonality formulations, we see a new
ingredient appearing in the SoV framework, namely lower-length
Baxter operators, which appear at each order in perturbation theory. This
suggests that the perturbative SoV framework for $\mathcal{N}=4$ super
Yang--Mills theory, even at rank one, should depend on \emph{all Q-functions,
as well as lower-length Baxter operators acting on them, with appropriate
transcendentality}.

\paragraph{3. Residues.}

To find Functional SoV orthogonality relations, one must obtain
as many measures as integrals of motion and conserved
charges appear in the equation.
In order to not introduce extra
terms in this counting, one usually searches for measures that have
vanishing residues. However, there are some previously overlooked
measures which do have residues, but still do not change this counting.
Namely, \emph{measures might have non-zero residues, as long as they
are proportional to differences of integrals of motions and
conserved charges, multiplied by bilinear integrals of Q-functions
with modified (state-independent) measures}.

\vspace{1em}

It would be nice to revisit known integrable models that have elusive
SoV representations, for example, the supersymmetric
$\alg{su}(m|n)$ spin chain, or the $\alg{su}(2)$ XXZ spin chain,
which so far admit separated wave functions
\cite{Maillet:2018bim,Maillet:2019ayx}, but whose correlation functions
remain out of reach. Perhaps in these cases, one also needs to relax
conditions on the measures and residues compared to the rational
$\alg{su}(n)$ set-up.

Although orthogonality seems to be the correct guideline to search for
SoV representations for two-point functions, there is no guarantee
that the measures we obtained here will be useful for three-point
functions. For example, in the $\alg{su}(2)$ sector, the measures must
be slightly modified when going from two-point to three-point
functions~\cite{Bercini:2022jxo}. The SoV-inspired proposal~\eqref{claim}
can account for this difference in a neat way: Due to the asymmetry in
$S$ and $J$, the determinant expressions
\begin{align}
   \det \mathcal{M}_{L,\ell}[Q_S,Q_J]
   \,,\qquad
   \det \mathcal{M}_{L,\ell}[Q_J,Q_S]
   \,,\qquad
   \sqrt{\det \mathcal{M}_{L,\ell}[Q_S,Q_J] \det\mathcal{M}_{L,\ell}[Q_S,Q_J]}
\end{align}
are all different, and while it is natural that their symmetric
combination could be used to compute two-point functions, the other
non-symmetric determinants could compute the overlaps with the vacuum
and three-point functions.

One might be worried that the unstable relative factor
between the SoV expressions and the Gaudin norm spoils such explorations,
but one hope is that these factors become simple when considering
three-point functions. This is exactly what was observed for the $\alg{sl}(2)$
and $\alg{su}(2)$ sectors: The normalization factors
between SoV expressions for the norm $\GaudinB$ and the SoV
overlaps with the vacuum (the so-called $\mathcal{A}$ of the
integrable hexagons framework~\cite{Basso:2015zoa}) were built from
complicated combinations of Bethe roots. On the contrary, the
normalization factor of the ratio of these two quantities, which is
precisely what evaluates three-point functions, was given by a simple
factorial~\cite{Bercini:2022jxo}.
Perhaps more about the relation
between these objects can be gained by studying the orbifolded or
twisted setups recently analyzed using
hexagons \cite{lePlat:2025eod}.

The all-loop expressions we proposed here could open a new way for
strong-coupling explorations in the Separation of Variables formalism.
For small operators, our expressions break down due to wrapping
corrections, and we have nothing to say about strong coupling. But for
parametrically large operators, our expressions should remain valid, and
perhaps there could be some interesting semi-classical limits emerging
in the SoV formalism, like the ones observed
in~\cite{Jiang:2015lda,Gromov:2011jh,Kostov:2014iva,Bettelheim:2014gma,Kazama:2013qsa}.

Another interesting possibility is to consider an analytic continuation
in the spin~$S$. The Baxter equation for twist three at complex spin was
recently extensively analyzed in~\cite{Homrich:2024nwc}. As was argued
in~\cite{Homrich:2022cfq}, one cannot simply upgrade the SoV
expression for the three-point function built from $Q^{(1)}_S$ by
promoting it to its entire analytic continuation \cite{Janik:2013nqa},
built from both $Q^{(1)}$ and $Q^{(2)}$ with $i$-periodic
coefficients. This may be related to the fact that in the QSC, there are
two distinct objects which become
$Q^{(1)}$ for integer $S$ at tree level, but which differ as soon as
$S$ is complex.%
\footnote{These are $Q_{12|12}$ discussed
in the main text, and another object denoted $\mu_{12}^+$.}
It would be nice to understand how these
analytic continuations fit in the SoV framework. Perhaps the relevant
objects for a complex $S$ SoV expression are neither $Q^{(1)}$ nor
$Q^{(2)}$, but precise linear combinations of these two Q-functions
that have yet to be found.

Another limit that would be nice to consider is the large-spin limit.
This limit is another mechanism that could suppress the wrapping corrections,
and still be valid for small-twist operators. Differently from the
integrable hexagons formalism~\cite{Basso:2015zoa}, where the spin
controls the number of partitions one needs to sum over, in the
SoV expressions the spin only controls the degree of the polynomial part
of the Q-functions. This makes SoV-type expressions amenable to both
numerical as well as analytical explorations at large spin. Better
understanding this limit could shine further light on the dualities
between Wilson loops and correlation functions that are present in
$\mathcal{N}=4$ super
Yang--Mills~\cite{Alday:2010zy,Alday:2010ku,Bercini:2020msp,Bercini:2021jti}.

We hope that our proposal, or at least the general lessons we are
suggesting, will help to improve the Separation of Variables
framework in $\mathcal{N}=4$ super Yang--Mills and other integrable
models, and bring us one step closer to the same level of computational
efficiency as the Quantum Spectral Curve.

\section*{Acknowledgments}
\label{acknow}

We are grateful to B.~Basso, J.~Br\"odel, S.~Ekhammar, N.~Gromov,
A.~Homrich, E.~Im, G.~Lefundes, A.~Pribytok, N.~Primi, D.~Serban,
R.~Tateo, P.~Vieira, and D.~Volin for useful discussions.
We are especially grateful to A.~Homrich and P.~Vieira for numerous
suggestions, and collaboration on several topics discussed here.
The work of T.\,B., C.\,B., D.\,L. and P.\,R. was funded by the Deutsche
Forschungsgemeinschaft (DFG, German Research Foundation) Grant No.
460391856.
T.\,B., C.\,B., D.\,L. and P.\,R. acknowledge support from DESY
(Hamburg, Germany), a member of the Helmholtz Association HGF,
and by the Deutsche
Forschungsgemeinschaft (DFG, German Research Foundation) under
Germany's Excellence Strategy -- EXC 2121 ``Quantum Universe'' --
390833306.
D.\,L. further thanks University of Turin and ETH Z\"urich for the generous
and repeated hospitality, Antonio Antunes for his invaluable insights into the
history of integrability and Lucia Bianco for her unique and
unwavering support. A.\,C. participates to the project
HORIZON-MSCA-2023-SE-01-101182937-HeI, and acknowledges support from
this action and from the INFN SFT specific initiative. He also thanks
DESY for the warm hospitality during two extended visits.

\appendix

\section{Bethe equations}
\label{appBaxter}

The Bethe roots $v_n$, $n=1,\dots,S$, are solution to the asymptotic
Bethe ansatz (ABA) equations~\cite{Beisert:2005fw,Arutyunov:2004vx}
\begin{equation}
    \left(\frac{x^{+}_k}{x^{-}_k}\right)^L = \prod_{j\neq k}^{S}\frac{x_k^{-}-x_j^{+}}{x_k^{+}-x_j^{-}}\frac{1-g^2/x_k^+ x_j^-}{1-g^2/x_k^- x_j^+}e^{2i\, \theta(v_k,v_j)}
    \,.
    \label{eq:BetheEq2}
\end{equation}
In addition to the ABA equations, one also needs to impose cyclicity
of the trace in the form of the zero-momentum condition
\begin{equation}
    \prod_{k=1}^S \frac{x^+_k}{x^-_k}=1\,,
\end{equation}
where the Zhukovsky variables are
\begin{equation}
    x_k^{\pm} = x\left(v_k \pm \frac{i}{2}\right) \quad \text{with} \quad x(u) = \frac{u+\sqrt{u^2-4g^2}}{2}\,.
\end{equation}
The object $\theta(u,v)$ is the so-called dressing phase \cite{Beisert:2004hm,
Arutyunov:2004vx,
Janik:2006dc,
Beisert:2006ib,
Beisert:2006ez,
Dorey:2007xn,Vieira:2010kb}. In weak-coupling perturbation theory, the phase
factor starts to contribute only at
order $\mathcal{O}(g^6)$, and is built out of several
infinite sums. At any finite order $\mathcal{O}(g^{2\Lambda})$,
these sums truncate, and the dressing phase can be written as
\begin{align}
    \theta(u,v) & = \sum_{r=1}^{\Lambda}\sum_{s=1}^{\Lambda}\sum_{n=0}^{\Lambda-r-s}g^{2(r+s+n)}\beta_{r,s,n}t_r(u)t_s(v)\,, \\
    \beta_{r,s,n} & =  2(-1)^n\frac{\sin(\frac{\pi}{2}(r-s))\zeta_{2n+r+s}\Gamma(2n+r+s)\Gamma(2n+r+s+1)}{\Gamma(n+1)\Gamma(n+r+1)\Gamma(n+s+1)\Gamma(n+r+s+1)}\,, \\
    t_r(u) & = \left(\frac{1}{x^{+}(u)}\right)^r-\left(\frac{1}{x^{-}(u)}\right)^r\,.
\end{align}

Using an extension of the following \mathematica code presented in~\cite{IntSchool},
\begin{lstlisting}[basicstyle={\small\ttfamily}]
ns[L_,S_] = Select[Subsets[Range[-(L+S-3)/2, (L+S+3)/2], {S}],
    Mod[Total[#], L] == 0 &];
BAE[L_, S_, n_] := Table[2 L ArcTan[2u[j]] + 2Sum[If[j==k,0,
    ArcTan[u[j] - u[k]]],{k,1,S}] - 2 Pi ns[L,s][[n,j]],{j,1,S}]
U[L_,S_] := Table[#[[2]] &/@ FindRoot[BAE[L,S,n], Table[{u[j], j-S/2},
    {j, S}], WorkingPrecision -> 150], {n,Length@ns[L, S]}]
\end{lstlisting}
it is easy to enumerate the $\alg{sl}(2)$
states for any twist $L$ and spin $S$, and then compute their
corresponding Bethe roots $v_n$ at leading order with arbitrary
numerical precision.
After finding the leading-order
solutions \lstinline!U[L,S]! for the Bethe
equations~\eqref{eq:BetheEq2}, it is easy
to compute their loop corrections by simply
linearizing the Bethe equations around these seed values.

In this way one can evaluate the Bethe roots at any order in
perturbation theory with arbitrary numerical precision. It is then
trivial to assemble
them to write the polynomial part of the Q-functions $\mathcal{P}(u) =
\prod_{k=1}^S(u-v_k)$, and to write the conserved
charges that appear in the dressing~\eqref{dressingExpand}, via
\begin{align}\label{eqn:chargesfromroots}
    q_n^{\pm}[Q_S]&=\sum_{k=1}^{S}\left(\left(\frac{i}{x_k^{+}}\right)^n\pm\left(\frac{-i}{x_k^{-}}\right)^n\right) \,.
\end{align}

Since the conserved charges are simple combinations of the Bethe
roots, it is also possible to write them as simple contour integrals
of the polynomial Q-functions
\begin{align}
    q_n^{+} &= \oint \frac{{\rm d}x}{4\pi i g^{n-1}}(-1)^n\left(1-\frac{g^2}{x^2}\right)\left(\frac{x^n}{g^n}-\frac{g^n}{x^n}\right)\left(i^n \frac{\mathcal{P}'(u+\tfrac{i}{2})}{\mathcal{P}(u+\tfrac{i}{2})} +(-i)^n\frac{\mathcal{P}'(u-\tfrac{i}{2})}{\mathcal{P}(u-\tfrac{i}{2})} \right) \,, \nn\\
    q_n^{-} &= \oint \frac{{\rm d}x}{4\pi i g^{n-1}}(-1)^n\left(1-\frac{g^2}{x^2}\right)\left(\frac{x^n}{g^n}-\frac{g^n}{x^n}\right)\left(i^{n-1} \frac{\mathcal{P}'(u+\tfrac{i}{2})}{\mathcal{P}(u+\tfrac{i}{2})} +(-i)^{n-1}\frac{\mathcal{P}'(u-\tfrac{i}{2})}{\mathcal{P}(u-\tfrac{i}{2})} \right)
    \,,
\label{eq:qContourIntegrals}
\end{align}
where the contour is the counter-clockwise unit circle in the
$x$-plane. The equivalence with \eqref{eqn:chargesfromroots} can be easily
checked by computing the integrals by residues.

\section{Gaudin Norm}
\label{appGaudin}

The Gaudin norm at leading order for a operator of twist-$L$ and spin $S$ is given by
\begin{equation}
  \GaudinB_{L,S} = \frac{\det(\partial_{v_i} \phi_j)}{\prod_{i \neq j}h(v_i,v_j)}
\end{equation}
where the derivatives act on the Bethe equations via
\begin{align}
    e^{i\phi_j} &= e^{i p(v_j) L} \prod_{k\neq j} S(v_j,v_k)
    \,,
\end{align}
and the momenta and S-matrices at leading order are given by
\begin{equation}
    e^{i p(u)} = \frac{u+\frac{i}{2}}{u+\frac{i}{2}}\,, \quad
    S(u,v) = \frac{h(u,v)}{h(v,u)}\, \quad \text{and} \quad
    h(u,v) = \frac{u-v}{u-v+i}\,.
\end{equation}

The relation between the enlarged orthogonality proposed
in~\eqref{claim} and the Gaudin norm at leading order for the first
two enlargings is
\begin{align}
  \mathcal{D}_{L,0}[Q_S,Q_J] &= \delta_{S,J}\times\frac{\Gamma(L)}{\Gamma(2S+L)} \left(Q_S\left(\sfrac{i}{2}\right)Q_S\left(-\sfrac{i}{2}\right)\right)^{-L} \times \GaudinB_L(S)
  \,,\nn\\
  \mathcal{D}_{L,1}[Q_S,Q_J] &= \delta_{S,J}\times\frac{\Gamma(L)}{\Gamma(2S+L)} \left(Q_S\left(\sfrac{i}{2}\right)Q_S\left(-\sfrac{i}{2}\right)\right)^{-(L+1)} \times \frac{\GaudinB_L(S)}{q_1^+(S)} \label{SoVtoB1L}
  \,.
\end{align}
For bigger enlargings, we can easily evaluate the proportionality
constant between these two quantities, see~\eqref{SoVtoBl}, but we were
not able to write it in a closed form in terms of the charges
like~\eqref{SoVtoB1L}.

\section{Evaluating Residues}
\label{appResidues}

There are two ways to compute the residues that are picked in the
contour manipulations of the initial equation~\eqref{leadingSelfAdj}. The first option
is to plug the on-shell values of the Q-functions (with all Bethe
roots and charges numerically computed), and then evaluate the residue
for many different states to check if it is zero, or if it can be
written as some combination of charges.

In this appendix we develop an alternative way, where we do not use
the explicit values of the Bethe roots and conserved charges for the
Q-functions. In this off-shell setup, we compute the residues as
analytic functions of the conserved charges. This trivializes the
process of finding measures with vanishing residues, and makes it easier
to find measures with residues that are proportional to conserved
charges. We exemplify this process by considering
N$^2$LO orthogonality for twist-two operators.

The residues that are picked in the contour manipulations that
transform the initial equation~\eqref{leadingSelfAdj} to the
final equation~\eqref{twoLoopSelfAdj} can be written explicitly
as
\begin{align}
  \texttt{res}_\mu = &\underset{u={i}/{2}}{\text{Res}}\left[\mathbb{Q}_S(u)\left(\brk{x^-}^2-\frac{g^4}{4}\brk1{q_2^{+}(S)+q_2^{+}(J)}\right)\mathbb{Q}_J(u-i)\mu(u)\right]+ \label{resExplicit}\\
  & \underset{u=-{i}/{2}}{\text{Res}}\left[\mathbb{Q}_S(u)\left(\brk{x^+}^2-\frac{g^4}{4}\brk1{q_2^{+}(S)+q_2^{+}(J)}\right)\mathbb{Q}_J(u+i)\mu(u)\right]+(S \leftrightarrow J)\,.\nonumber
\end{align}
Computing this residue for arbitrary measures can be done in three simple steps.

\paragraph{Step 1.}

Consider an ansatz for the measure. It can either
be in terms of hyperbolic functions like~\eqref{measureAnsatz}
\begin{equation}
  \mu(u) = \frac{\pi/2}{\cosh^2(\pi u)}(1+g^2(a_0 +a_1\tanh^2(\pi u))+\mathcal{O}(g^4))
  \label{appMeasure}
\end{equation}
Or, more generally, it can also be the Laurent series of the measure
around the poles $u=\pm i/2$, which is the only relevant part to
compute the residues
\begin{equation}
  \mu(u) = \sum_{n=-\Lambda_0}^{\Lambda_0}\left(\frac{a_n}{\left(u+\frac{i}{2}\right)^n}+\frac{a_n}{\left(u-\frac{i}{2}\right)^n}\right)+g^2\sum_{n=-\Lambda_1}^{\Lambda_1}\left(\frac{b_n}{\left(u+\frac{i}{2}\right)^n}+\frac{b_n}{\left(u-\frac{i}{2}\right)^n}\right)+\mathcal{O}(g^4)
  \label{LaurentSeries}
\end{equation}
where $\Lambda_i$ are cutoffs on the expansion whose values we explain
below.

\paragraph{Step 2.}

Plug the Laurent series~\eqref{LaurentSeries} and
the Q-functions~\eqref{QFunction} in the residue
expression~\eqref{resExplicit}. This can be done off-shell, namely we
do not need to specify the polynomial part of the Q-functions, and
neither the values of the charges appearing in the dressing
$e^{\alpha\cdot\sigma(u)}$. At N$^2$LO, it is sufficient to write:
\begin{equation}
  \mathbb{Q}_{S}(u) = \mathcal{P}_{S}(u)\left(1+g^2 q_1^{+}\psi_0^{+}-\frac{g^4}{2}\left(q_1^{+}\psi_2^{+}+q_2^{-}\psi_1^{-}\right)\right)+\mathcal{O}(g^6)
  \,.
  \label{offShellQ}
\end{equation}
Since $\mathcal{P}_{S}(u)$ is regular at
$u=\pm i/2$, all poles and zeros of the Q-functions at $u=\pm i/2$
are implemented via the Laurent series of the measure and the
off-shell Q-functions, so one can just evaluate the
residues~\eqref{resExplicit} explicitly.

For example, using the measure~\eqref{appMeasure} and the off-shell
Q-functions~\eqref{offShellQ}, we can write the
residue~\eqref{resExplicit} at $\text{NLO}$ as
\begin{align}
  \texttt{res}_\mu &= 0 + i g^2 \mathcal{P}_S\left(\sfrac{i}{2}\right)\mathcal{P}_J\left(\sfrac{i}{2}\right)(q_1^{+}(S)-q_1^{+}(J)) +\nonumber\\
  &+ 2g^2(2-a_1)\left(\mathcal{P}_S\left(\sfrac{i}{2}\right)\mathcal{P}^\prime_J\left(\sfrac{i}{2}\right)-\mathcal{P}_J\left(\sfrac{i}{2}\right)\mathcal{P}^\prime_S\left(\sfrac{i}{2}\right)\right)+\mathcal{O}(g^4)
  \,.
  \label{resExp2}
\end{align}

\paragraph{Step 3.}

At this stage, the evaluated residues depend on the
conserved charges $q_i^\pm$, the coefficients of the measure, and the
polynomial $\mathcal{P}_\text{S}(u)$ and its derivatives evaluated at
$u=\pm i/2$. To make this expression useful, we can use the fact that
these polynomial (and their derivatives) at specific values of the spectral
parameter are themselves written in terms of conserved charges. For
example, fixing the normalization of the Q-functions so that
$\mathcal{P}_\text{S}({i}/{2}) = 1$, we then have
\begin{align}
  \mathcal{P}_\text{S}^\prime\left(\sfrac{i}{2}\right) &= \frac{q_1^{+}}{2i}+g^2\frac{q_3^{+}}{2i}+g^4\frac{q_5^{+}}{2i}+\mathcal{O}(g^6) \\
  \mathcal{P}_\text{S}^{\prime\prime}\left(\sfrac{i}{2}\right) &= \left(\frac{q_2^{+}}{2}-\frac{(q_1^{+})^2}{4}\right)+\frac{g^2}{2}\left(q_4^{+}-q_1^{+}q_3^{+}\right)+\frac{g^4}{2}\left(3q_6^{+}-q_1^{+}q_5^{+}-\frac{(q_3^{+})^2}{2}\right)+\mathcal{O}(g^6)
\end{align}
The outcome of implementing these three steps is that for any given
cut-offs $\Lambda_i$ (or combination of hyperbolic functions) we can
compute the residue~\eqref{resExplicit} as a function of the conserved
charges of the Q-functions and the coefficients in the measure ansatz. In
the case of~\eqref{resExp2}, the residue becomes
\begin{equation}
  \texttt{res}_\mu = 0 - i g^2(a_1-3)(q_1^{+}( {S})-q_1^{+}( {J}))+\mathcal{O}(g^4) \,.
\end{equation}
This in turns fixes the measure at one-loop order to $a_1=3$.
We conclude that for the dressing parameter
$\alpha={1}/{2}$, at NLO there is a unique measure that has vanishing
residues.%
\footnote{The remaining free coefficient $a_1$ does not affect the
orthogonality of different Q-functions, it only enters the norm of
equal Q-functions.} This statement remains true at the next order, and the resulting
unique N$^2$LO measure is~\eqref{mu1}. The same analysis can be done for arbitrary values
of $\alpha$, and at any loop order. By doing precisely such
computations, we obtain that no two measures with vanishing residues
exist for twist-two operators.

In principle, one could try to search for alternative measures by
considering an ansatz with arbitrary large values of the cut-offs
$\Lambda_i$ (or arbitrarily large powers of $\tanh(\pi u)$ in the
ansatz~\eqref{measureAnsatz}). We tried to do that, but it
seems unlikely that large values of these cut-offs will result in
vanishing residues. Since they control the degrees of the poles, they
will result in residues with higher derivatives of the Q-functions,
which in turn will generate higher conserved charges in the residues,
that themselves need to be canceled.

\bibliographystyle{nb.bst}
\bibliography{refs.bib}
\end{document}